\documentclass[twocolumn]{aa}
\usepackage{amsmath,amssymb,amsfonts,bm,mathtools}
\usepackage{physics}
\usepackage{graphicx}
\usepackage{xcolor}
\usepackage[normalem]{ulem}
\usepackage{txfonts}
\usepackage{placeins}
\usepackage{threeparttable}
\usepackage{orcidlink}
\usepackage{array}
\usepackage{hyperref}
\usepackage{tabularx}
\hypersetup{
    colorlinks,
    citecolor=blue,
    linkcolor=blue,
    urlcolor=blue
}
\newcommand{\br}{\boldsymbol{r}}

\newcommand{\bx}{\boldsymbol{x}}

\newcommand{\bOmega}{\boldsymbol{\Omega}}

\newcommand{\ii}{\mathrm{i}}
\newcommand{\ee}{\mathrm{e}}

\newcommand{\xx}{\bm{x}}
\newcommand{\XX}{\bm{X}}

\newcommand{\rr}{\bm{r}}
\newcommand{\RR}{\bm{R}}
\newcommand{\rp}{\bm{r}_{\rm p}}

\newcommand{\M}{\mathcal{M}}

\newcommand{\eplus}{\hat{\bm e}_{+}}

\newcommand{\zhat}{\hat{\bm z}}
\newcommand{\xhat}{\hat{\bm x}}
\newcommand{\yhat}{\hat{\bm y}}

\newcommand{\rhat}{\hat{\bm r}}

\newcommand{\R}{\mathcal{R}}

\newcommand{\ab}{a_{\rm bin}}

\begin{document}
\title{Dynamical friction in stratified stellar envelopes}
\author{
    Damien Gagnier\corrauth{damien.gagnier@uni-heidelberg.de}\orcidlink{0000-0002-1904-2740}
}
\institute{
    Zentrum für Astronomie der Universität Heidelberg, Astronomisches Rechen-Institut,
    Mönchhofstr. 12-14, D-69120 Heidelberg, Germany\\
    Heidelberger Institut für Theoretische Studien,
    Schloss-Wolfsbrunnenweg 35, 69118 Heidelberg, Germany
}
\abstract
{Dynamical friction prescriptions used for common-envelope and planetary engulfment inspirals often assume a  homogeneous medium and/or rectilinear perturber motion. A gravitating object embedded in a giant-star envelope instead excites an orbit-scale wake while moving on a curved orbit through a finite, radially stratified medium.}
{We quantify the linear barotropic acoustic wake and the associated gravitational back-reaction for low-mass perturbers on circular orbits.}
{We formulate the barotropic acoustic response of a weak point perturber on a circular orbit in a hydrostatic, spherically stratified gaseous medium. The enthalpy perturbation is expanded in spherical harmonics and Fourier modes. The force is written as an adjacent-multipole sum, with coefficients computed from the retarded acoustic Green function. We apply the formulation to single perturbers in power-law density profiles and giant-star envelope models, and to double perturbers in power-law density backgrounds.}
{We find that stratification affects dynamical friction through the global structure of the wake, not only through the local density, sound speed, and Mach number at the perturber position. The radial component is set  by the low-order, orbit-scale wake and can strongly differ in amplitude and sign from the homogeneous-medium result. The azimuthal component is also modified by stratification, but in the supersonic regime it retains the Coulomb-logarithmic sensitivity of the homogeneous problem.
In double-perturber systems, the companion wake can substantially change the radial force and reduce the azimuthal drag on a given component, but, unlike the perturber's own wake, it has no local Coulomb-logarithmic contribution. For the adopted giant-star envelope profiles, the azimuthal drag exerted by the stratified wake gives shorter inspiral times than uniform-medium prescriptions evaluated with the same local background quantities.}
{The gravitational back-reaction of the linear barotropic wake in stratified stellar envelopes combines a global acoustic response with a cutoff-sensitive drag contribution.
The formulation provides a flexible tool for computing embedded-perturber wakes in prescribed radial stratifications and is a first step toward computationally efficient, self-consistent models of common-envelope and planetary-engulfment inspirals.}

\maketitle
\nolinenumbers

\section{Introduction}

The inspiral of a gravitating object inside a giant-star envelope is driven by the exchange of orbital energy and angular momentum with the surrounding gas. A major channel for this exchange is the gaseous wake raised by the perturber, whose gravitational back-reaction exerts a drag force commonly referred to as dynamical friction. This force influences the inspiral rate, the envelope response, and the fate of the embedded object, making it a key mechanism in common-envelope evolution \citep{Paczynski1976} and planetary engulfment \citep{Alexander1967}.

In collisionless systems, the local theory for dynamical friction was introduced by \citet{Chandrasekhar1943} and later generalized to the collective and resonant response of finite, inhomogeneous systems \citep[e.g.,][]{Kalnajs1971,Mulder1983,TW84,Weinberg1986}. In gaseous media, the analogue force is the back-reaction of a compressible wake, whose structure depends on sound-wave propagation, the perturber Mach number, and the geometry of the trajectory \citep[e.g.,][]{Dokuchaev1964,Ruderman1971,Rephaeli1980,Ostriker1999,Sanchez2001,kim2007,Desjacques2022,ONeill2024,Eytan2024,Eytan2026}.

Dynamical friction  prescriptions are useful closures for binary stellar evolution and planetary migration calculations. For example, \citet{RomanGarza2026} and \citet{Alidib2026} use the rectilinear homogeneous medium formula of \citet{Ostriker1999}, \citet{Fragos2019} combine the formula of \citet{Ostriker1999} with a fitting formula calibrated from the local hydrodynamic simulations of \citet{MacLeod2015}, while \citet{Bronner2024} adopt prescriptions based on \citet{kim2007} and \citet{kim2010}. However, such prescriptions have important limitations for modelling inspirals through stellar envelopes. Homogeneous-medium prescriptions \citep[e.g.,][]{Ostriker1999,kim2007,Desjacques2022} neglect the radial stratification of the stellar envelope. Local simulations centered on the embedded object \citep[e.g.,][]{MacLeod2015,MM2017,De2020,Prust2024,Gagnier2026} resolve the nonlinear near field and part of the surrounding wake. They however only cover a region spanning a few accretion radii around the perturber, and therefore do not capture the full orbit-scale wake response. Furthermore, prescriptions based on rectilinear motion \citep{Ostriker1999} omit the radial force toward the centre of motion that arises from orbital curvature \citep[e.g.,][]{kim2007,Desjacques2022}. A spiralling-in perturber instead excites a global wake while moving on a curved orbit through a finite, radially stratified medium. The resulting force may therefore depend on the envelope structure away from the perturber, not only on the density, sound speed, and Mach number evaluated at its position.

The circular-orbit acoustic problem has been solved analytically in two important limits. \citet{Desjacques2022} obtained the steady linear response of a homogeneous medium, and, more recently, \citet{Eytan2024} extended the calculation to the singular isothermal sphere (SIS), showing that a spherical stratification can substantially modify the acoustic response.

In this paper, we calculate the linear barotropic acoustic wake of a weak perturber on a circular orbit in a hydrostatic, spherically stratified gaseous medium. The calculation generalizes the homogeneous solution of \citet{Desjacques2022} and the SIS solution of \citet{Eytan2024} to general hydrostatic radial stratification. We also extend  the formalism to double-perturber systems. The physical setup and force decomposition are introduced in Sect.~\ref{sec:setup}, followed by the derivation of the acoustic response in Sect.~\ref{sec:approx} and its uniform-medium limit in Sect.~\ref{sec:nostrat}. We then apply the formalism to uniform media, power-law density profiles, and stellar-envelope models in Sect.~\ref{sec:applications}, before discussing its assumptions, limitations, and implications in Sect.~\ref{sec:discussion}. We conclude in Sect.~\ref{sec:conclu}.

\section{Physical setup}
\label{sec:setup}
The symbols used throughout the paper are summarized in Table~\ref{tab:notation}.
We consider a point perturber of mass $M$ moving on a circular orbit of radius $a$ and angular velocity $\bOmega = \Omega \zhat$. Its instantaneous position reads
\begin{equation}
  \rp(t)=a\left[\cos(\Omega t)\xhat+\sin(\Omega t)\yhat\right].
  \label{eq:orbit_position}
\end{equation}
The background medium is taken to be stationary and spherically symmetric with density $\rho_0(r)$, pressure $p_0(r)$, and sound speed $c_s(r)$. The orbital Mach number is defined as
\begin{equation}
  \M \equiv \frac{\Omega a}{c_s(a)}.
  \label{eq:orbital_mach}
\end{equation}
We write $\rho = \rho_0 + \rho_1$, $p = p_0 + p_1$ where $\rho_1$ and $p_1$ are the wake density and pressure perturbation, respectively. The gravitational force exerted by the wake on the perturber reads
\begin{equation}
  \bm F(t)=GM\int \dd^3 x \rho_1(\xx,t) \frac{\xx-\rp(t)}{\|\xx-\rp(t)\|^3},
  \label{eq:force_general}
\end{equation}
and can be decomposed into a radial component $F_r$, a tangential component $F_\varphi$, and a vertical component $F_z$ normal to the orbital plane. For a circular orbit in the $z=0$ plane with a spherically symmetric background, reflection symmetry across that plane gives $F_z = 0$. Because the background is spherically symmetric and the forcing is periodic, the steady retarded wake is time independent in the frame co-rotating with the perturber.  We choose $t=0$ as a reference phase, for which $\br_{\rm p}=a\xhat$, so that the radial and azimuthal components are respectively aligned with $\xhat$ and $\yhat$. We combine the in-plane force into a single complex quantity using the helicity basis\footnote{The definition of $\hat{\bm e}_{\pm}$ varies in the literature, e.g.\ \citet{Varshalovich1988} use $\hat{\bm e}_{\pm}=\mp(\xhat+\ii\yhat)/\sqrt{2}$, while \citet{Desjacques2022} use $\hat{\bm e}_{\pm}=(\ii\yhat \mp \xhat)/\sqrt{2}$. The choice affects expressions in intermediate steps but not the final result.}
\begin{equation}
  \mathcal F_+ \equiv  \eplus\cdot\bm F(0)
  =\frac{F_r+\ii F_\varphi}{\sqrt2}, \quad \eplus=\frac{\xhat+\ii\yhat}{\sqrt2}.
  \label{eq:Fplus_def}
\end{equation}
Hence,
\begin{equation}
  F_r=\sqrt2\Re\left(\mathcal F_+\right),
  \quad
  F_\varphi=\sqrt2\Im\left(\mathcal F_+\right).
  \label{eq:FrFp}
\end{equation}

\section{The linear barotropic acoustic response}\label{sec:approx}
We work in the linear barotropic acoustic approximation in which the pressure and density perturbations satisfy,
\begin{equation}
  p_1 = c_s^2(r)\rho_1,
\end{equation}
so that the thermodynamics enters only through the prescribed sound speed. This approximation retains the compressive acoustic response of the gas, but excludes the non-barotropic buoyancy perturbations associated with entropy advection. We also neglect rotation, convection, accretion, and the self-gravity of the perturbation. Its regime of validity and possible extensions are discussed in Sect.~\ref{sec:discussion}.

\subsection{Forced acoustic wave equation}
Introducing the enthalpy perturbation
\begin{equation}
  h(\xx,t)\equiv \frac{p_1(\xx,t)}{\rho_0(r)},
  \label{eq:h_def}
\end{equation}
and noting that the unperturbed medium is time independent, the linearized mass-conservation equation can be written as
\begin{equation}
\frac{\rho_0(r)}{c_s^2(r)}\pdv{h}{t} +\nabla\cdot\left[\rho_0(r)\bm v\right]=0 .
\label{eq:strat_cont_h}
\end{equation}
The linearized Euler equation is
\begin{equation}
\pdv{\bm v}{t}=-\nabla h-\nabla\Phi_p,
\label{eq:strat_mom}
\end{equation}
where $\Phi_p(\xx,t)=-GM/\|\xx-\rp(t)\|$
is the perturber's gravitational potential. For a barotropic hydrostatic background, the force associated with the background gravitational field can be absorbed into the enthalpy gradient. Taking the time derivative of Eq.~\eqref{eq:strat_cont_h} and substituting Eq.~\eqref{eq:strat_mom} yields the forced acoustic wave equation
\begin{equation}
	\frac{\rho_0}{c_s^2}\pdv[2]{h}{t} - \nabla\cdot\left(\rho_0\nabla h\right)  = \nabla\cdot\left(\rho_0\nabla\Phi_p\right).
  \label{eq:barotropic_acoustic_wave}
\end{equation}
We expand the enthalpy perturbation and the perturber potential in spherical harmonics,
\begin{equation}
  h(\xx,t)=\sum_{\ell m}h_{\ell m}(r,t)Y_\ell^m(\rhat), \quad \Phi_p(\xx,t)=\sum_{\ell m}\Phi_{\ell m}(r,t)Y_\ell^m(\rhat),
  \label{eq:harmonic_expansion}
\end{equation}
and we decompose the time dependence of each spherical-harmonic coefficient into Fourier modes
\begin{equation}
  h_{\ell m}(r,t)   =  \int\frac{\dd\omega}{2\pi} h_{\ell m}(r;\omega)\ee^{-\ii\omega t},
  \quad \Phi_{\ell m}(r,t) = \int\frac{\dd\omega}{2\pi} \Phi_{\ell m}(r;\omega)\ee^{-\ii\omega t}.
  \label{eq:fourier_decomposition}
\end{equation}
With this convention, $\partial_t^2\rightarrow -\omega^2$, and each $(\ell,m,\omega)$ component of Eq.~\eqref{eq:barotropic_acoustic_wave} satisfies
\begin{equation}
  \left[\mathcal L_\ell-\omega^2\frac{\rho_0}{c_s^2}\right] h_{\ell m}(r;\omega) =  -\mathcal L_\ell\Phi_{\ell m}(r;\omega),
\end{equation}
where the radial operator is
\begin{equation}
  \mathcal L_\ell f   =  -\frac{1}{r^2}\frac{\dd}{\dd r}  \left(r^2\rho_0\frac{\dd f}{\dd r}\right) +\rho_0\frac{\ell(\ell+1)}{r^2}f .
  \label{eq:Lell_def}
\end{equation}
The retarded radial Green function is defined by
\begin{equation}
  \left[\mathcal L_\ell-\omega^2\frac{\rho_0}{c_s^2}\right] g_\ell(r,r';\omega)  =\frac{\delta(r-r')}{r^2},
  \label{eq:Green_strat}
\end{equation}
and the solution reads
\begin{equation}
  h_{\ell m}(r;\omega) =-\int_0^\infty \dd r' r'^2  g_\ell(r,r';\omega)\mathcal L_\ell\Phi_{\ell m}(r';\omega).
\end{equation}

\subsection{Dynamical friction}\label{sec:DF}
The gravitational force exerted by the wake defined in Eq.~\eqref{eq:force_general} can be rewritten as
\begin{equation}
  \bm F(t)=-M\nabla_{\RR}\Psi(\RR,t)\big|_{\RR=\rp(t)} ,
  \label{eq:force_from_wake_potential}
\end{equation}
where we introduced the gravitational potential of the wake,
\begin{equation}
  \Psi(\RR,t) \equiv -G\int \dd^3x\,\frac{\rho_1(\xx,t)}{\|\RR-\xx\|}.
  \label{eq:wake_potential_def}
\end{equation}
We expand the wake potential in spherical harmonics, 
\begin{equation}
  \Psi(\RR,t)=\sum_{\ell m}\Psi_{\ell m}(R,t)Y_\ell^m(\hat\RR).
\end{equation}
Using the standard Laplace expansion 
\begin{equation}
	\frac{1}{\|\RR-\xx\|} = \sum_{\ell m}\frac{4\pi}{2\ell+1}q_\ell(r,R) Y_\ell^m(\hat\RR)Y_\ell^{m*}(\xhat),
  \label{eq:laplace_expansion}
\end{equation}
where $q_\ell(r,R)\equiv r_<^\ell/r_>^{\ell+1}$, $r_<\equiv\min(r,R)$, and $r_>\equiv\max(r,R)$, the radial coefficient can be written
\begin{equation}
\Psi_{\ell m}(R,t)=\psi_{\ell m}(R;m\Omega)\ee^{-\ii m\Omega t},
  \label{eq:wake_potential_time_dependence}
\end{equation}
with
\begin{equation}
\psi_{\ell m}(R;\omega)=-G\frac{4\pi}{2\ell+1}\int_0^\infty \dd r r^2\frac{\rho_0(r)}{c_s^2(r)}q_\ell(r,R)H_{\ell m}(r;\omega) ,
  \label{eq:wake_potential_amplitude}
\end{equation}
and
\begin{equation}
H_{\ell m}(r;\omega) \equiv -\int_0^\infty \dd r' r'^2 g_\ell(r,r';\omega) \mathcal L_\ell\Phi_{\ell m}^{(0)}(r') .
\label{eq:Hlm_def_general}
\end{equation}
Here $\Phi_{\ell m}^{(0)}(r)$ is the radial amplitude of the perturber potential, excluding the factor $\ee^{-\ii m\Omega t}$:
\begin{equation}
\Phi_{\ell m}^{(0)}(r)= -GM\frac{4\pi}{2\ell+1}A_{\ell m}q_\ell(r,a),
\label{eq:Phi0_q_def}
\end{equation}
where 
\begin{equation}
  A_{\ell m}\equiv Y_\ell^{m*}\!\left(\frac{\pi}{2},0\right) =\left[\frac{2\ell+1}{4\pi}\frac{(\ell-m)!}{(\ell+m)!}\right]^{1/2}P_\ell^m(0) 
  \label{eq:Alm_def}
\end{equation}
in the Condon--Shortley phase convention \citep[e.g.,][]{Arfken1995}. Finally, we evaluate the helicity component of the force, noting that, at an arbitrary point on the orbital plane $(R,\pi/2,\varphi)$,
\begin{equation}
\eplus\cdot\nabla =\frac{\ee^{\ii\varphi}}{\sqrt2} \left(\partial_R+\frac{\ii}{R}\partial_\varphi\right) .
\end{equation}
We obtain
\begin{equation}
	F_+(t)=\ee^{\ii\Omega t}\mathcal{F}_+ ,
\end{equation}
with 
\begin{equation}
	\mathcal{F}_+ =\frac{G^2M^2}{\sqrt2} \sum_{\ell=0}^{\infty}\sum_{m=-\ell}^{\ell} \left(\frac{4\pi}{2\ell+1}\right)^2 A_{\ell m}^2 \mathfrak F_{\ell m}(a;m\Omega), 
\label{eq:Fplus_strat_matrix}
\end{equation}
where
\begin{align}
   \mathfrak F_{\ell m}(a;m\Omega)  &\equiv \int_0^\infty \dd r\,r^2 w(r)\left[\partial_R q_\ell(r,R)-\frac{m}{R}q_\ell(r,R)\right]_{R=a} \mathcal Y_{\ell m}(r;\omega) \label{eq:Frak_lm_def} \\ 
    &= (\ell-m)\mathcal J_{\ell,-}(m) - (\ell+1+m)\mathcal J_{\ell,+}(m),
\end{align}
\begin{equation}
  \mathcal J_{\ell,+}(m) \equiv \int_0^a \dd r r^2w(r) \frac{r^\ell}{a^{\ell+2}} \mathcal Y_{\ell m}(r;m\Omega),
  \label{eq:Jell_plus}
\end{equation}
\begin{equation}
  \mathcal J_{\ell,-}(m) \equiv \int_a^\infty \dd r\,r^2w(r) \frac{a^{\ell-1}}{r^{\ell+1}}\mathcal Y_{\ell m}(r;m\Omega),
  \label{eq:Jell_minus}
\end{equation}
\begin{equation}
  \mathcal Y_{\ell m}(r;\omega) \equiv \int_0^\infty \dd r' r'^2   g_\ell(r,r';\omega)\,\mathcal L_\ell q_\ell(r',a),
  \label{eq:Ylm_response_kernel}
\end{equation}
and $w(r)\equiv \rho_0(r)/c_s^2(r)$. Finally, letting $s=\pm1$, defining
\begin{equation}
  D_\ell  \equiv  \left(\frac{4\pi}{2\ell+1}\right)^2 (2\ell+1) \left(\frac{8\pi}{3}\right)^{1/2},
  \label{eq:Dell_def}
\end{equation}
and using the helicity recoupling coefficients $\mathcal H_s(\ell,m)$ defined in Eq.~\eqref{eq:Hs_single_closed}, the force may be written in the equivalent adjacent-multipole form
\begin{equation}
\begin{aligned}
  \mathcal F_+  &=  \frac{G^2M^2}{\sqrt2}  \sum_{\ell m}\sum_{s=\pm1}  s D_\ell  A_{\ell m}A_{\ell+s,m+1} \mathcal H_s(\ell,m) \mathcal J_{\ell,s}(m) \\ 
  & =   \frac{G^2M^2}{\sqrt2} \sum_{\ell=0}^{\infty} \sum_{s=\pm1}  \sum_{n\ge1}  s A_{\ell,n-1}A_{\ell+s,n}\mathcal H_s(\ell,n-1) \\ 
  & \times \left[  D_\ell \mathcal J_{\ell,s}(n-1) -  D_{\ell+s}\,\mathcal J_{\ell+s,-s}(n)^* \right].
\end{aligned}
  \label{eq:Fplus_strat_adjacent_unpaired}
\end{equation}
This is the stratified analogue of the homogeneous medium expression of \citet{Desjacques2022}. All dependence on the background stratification is contained in $w(r)$, in the operator $\mathcal L_\ell$, and in the retarded Green function $g_\ell$. The physical force components are
\begin{equation}
	F_r(t)=\sqrt2 \Re\!\left[F_+(t)\right],
  \quad
  F_\varphi(t)=\sqrt2 \Im\!\left[F_+(t)\right].
\end{equation}
At $t=0$, these reduce to the real and imaginary parts of $\mathcal{F}_+$ (see Eq.~\eqref{eq:FrFp}). The zero-frequency mode requires separate treatment. Setting $m=0$ removes the acoustic term from the radial equation, giving
\begin{equation}
	{\mathcal L}_\ell {\mathcal Y}_{\ell 0} (r) = {\mathcal L}_\ell q_\ell(r, a).
\end{equation}
For $\ell > 0$, regularity at the origin and absence of infinitely growing multipole yield
\begin{equation}
        \mathcal Y_{\ell 0} (r) = q_\ell(r, a).
\end{equation}
For $\ell = 0$, however, $\mathcal L_0 C = 0$ for any constant C, and hence 
\begin{equation}
        \mathcal Y_{0 0} (r) = q_0(r, a) + C.
  \label{eq:static_monopole_general}
\end{equation}
The constant $C$ fixes the global monopole convention. The choice $C=0$ corresponds to a decaying enthalpy perturbation at infinity and is the  convention appropriate for the homogeneous-medium limit. The choice $C = -1/a$ sets the monopole enthalpy perturbation to zero inside the orbit, and is the convention used to reproduce the results of \citet{Eytan2024} (see Sect.~\ref{sec:applications}). This choice is not appropriate to the homogeneous case because it would impose a nonzero density perturbation at infinity ($\rho_1 \to -GM \rho_0 /(a c_s^2))$.  For finite gas reservoirs such as stellar envelopes, we instead impose the mass-conserving constraint
\begin{equation}
	C =- \frac{\displaystyle\int_{r_{\min}}^{r_{\max}}  \dd r r^2 (\rho_0(r)/c_s^2(r))q_0(r,a)} {\displaystyle\int_{r_{\min}}^{r_{\max}} \dd r r^2 ( \rho_0(r)/c_s^2(r))}.
  \label{eq:static_monopole_mass_conserving_constant}
\end{equation}
Details of the numerical evaluation of Eq.~\eqref{eq:Fplus_strat_adjacent_unpaired} are given in Appendix~\ref{app:numerical_implementation}.

\section{The analytic uniform medium limit}
\label{sec:nostrat}
Here, we investigate the special case of a uniform gaseous medium. This case provides a consistency check against the circular-orbit result for a homogeneous medium of \citet{Desjacques2022}, as well as an analytic benchmark for the general stratified case presented in Sect.~\ref{sec:approx}. For constant $\rho_0$ and $c_s$, the radial operator Eq.~\eqref{eq:Lell_def} becomes
\begin{equation}
  \mathcal L_\ell f   =-\rho_0\left[\frac{1}{r^2}\frac{\dd}{\dd r}\left(r^2\frac{\dd f}{\dd r}\right) -\frac{\ell(\ell+1)}{r^2}f \right]
  \equiv -\rho_0 { \mathcal D_\ell } f ,
\end{equation}
where
\begin{equation}
  { \mathcal D_\ell } f\equiv \frac{1}{r^2}\frac{\dd}{\dd r}\left(r^2\frac{\dd f}{\dd r}\right)
  -\frac{\ell(\ell+1)}{r^2}f
  \label{eq:Dell_uniform}
\end{equation}
is the spherical radial Laplacian. The retarded radial Green function equation Eq.~\eqref{eq:Green_strat} becomes
 \begin{equation}
  \left[{ \mathcal D_\ell }+k^2\right]g_\ell(r,r')  =-\frac{1}{\rho_0}\frac{\delta(r-r')}{r^2} ,
  \label{eq:G_uniform_equation_D_new}
\end{equation}
with $k \equiv \omega/c_s$.
\subsection{The Bessel--Hankel Green function}\label{sec:bessel_hankel}
Away from the source point, $r\neq r'$, Eq.~\eqref{eq:G_uniform_equation_D_new} reduces to the spherical Bessel equation.
Imposing regularity of the solution as $r \to 0$, the condition that the solution behaves as an outgoing spherical wave as $r \to \infty$, and using the invariance property of the Green function under the change $r \leftrightarrow r'$ yields\footnote{If the opposite Fourier convention is used ($\propto e^{+\ii \omega t}$), the same retarded Green function is written with the conjugate spherical Hankel function $h_\ell^{(2)}(k r_>)$.} \citep{Joachain1975,Desjacques2022,Haber2024}
\begin{equation}
	g_\ell^{\rm uni}(r,r') =C_\ell\,j_\ell(k r_<)h_\ell^{(1)}(k r_>),
  \label{eq:G_uniform_ansatz_new}
\end{equation}
where $j_\ell(z)$ and $h_\ell^{(1)}(z)$ are respectively the spherical Bessel and spherical Hankel function of the first kind of order $\ell$ and argument $z$. The constant $C_\ell$ is set by integrating Eq.~\eqref{eq:G_uniform_equation_D_new} from $r = r'- \epsilon $ to  $r = r'+ \epsilon $, where $\epsilon$ is a positive infinitesimal quantity. It follows that $C_\ell = \ii k_m /\rho_0$, where $k_m = m\Omega/c_s$. Equation \eqref{eq:G_uniform_ansatz_new} can also be formulated in a Fourier--Bessel form \citep[see][]{Joachain1975,Haber2024}
\begin{equation}
 g_\ell^{\rm uni}(r,r') = \frac{2}{\pi\rho_0} \int_0^\infty\dd K\, \frac{K^2j_\ell(Kr)j_\ell(Kr')}{K^2-(k_m+\ii\epsilon)^2}.
  \label{eq:G_spectral_uniform_new}
\end{equation}
\subsection{The enthalpy response}
\label{sec:enthalpy_response}
We note that $q_\ell$ satisfies the radial Laplace equation (${ \mathcal D_\ell } q_\ell=0$) for $r \ne a$. At  $r = a$, $q_\ell$ is continuous, but its derivative is not. In particular
\begin{equation}
  \partial_rq_\ell\big|_{a_-}=\frac{\ell}{a^2}, \quad \partial_rq_\ell\big|_{a_+}=-\frac{\ell+1}{a^2} .
\end{equation}
Hence, we have
\begin{equation}
  \left[r^2\partial_rq_\ell\right]_{a_-}^{a_+}=-(2\ell+1),
\end{equation}
and thus
\begin{equation}
  \mathcal L_\ell q_\ell(r,a) =\rho_0\frac{2\ell+1}{a^2}\delta(r-a).
  \label{eq:Lq_uniform_delta_new}
\end{equation}
Injecting Eqs~\eqref{eq:G_uniform_ansatz_new} and \eqref{eq:Lq_uniform_delta_new} into the general stratified response Eq.~\eqref{eq:Hlm_def_general} yields
\begin{equation}
\begin{aligned}
  H_{\ell m}^{\rm uni}(r;m\Omega)
  &=GM\frac{4\pi}{2\ell+1}A_{\ell m}\int_0^\infty\dd r'\,r'^2 g_\ell^{\rm uni}(r,r'; m \Omega) \mathcal L_\ell q_\ell(r',a) \\
  &=4\pi GM A_{\ell m}\,\ii k_m j_\ell(k_m r_<)h_\ell^{(1)}(k_m r_>),
  \label{eq:H_uniform_limit_new}
\end{aligned}
\end{equation}
where we relabelled  $r_<\equiv\min(r,a)$ and $r_>\equiv\max(r,a)$.

\subsection{Dynamical friction}
We now inject Eq.~\eqref{eq:H_uniform_limit_new} into the general force formula Eq.~\eqref{eq:Fplus_strat_matrix}:
\begin{equation}
\begin{aligned}
  \mathcal F_{+}^{\rm uni} &=\frac{(4\pi)^2G^2M^2\rho_0}{\sqrt2 c_s^2} \sum_{\ell m}\frac{A_{\ell m}^2}{2\ell+1} \ii k_m
  \\ &\times \int_0^\infty\dd r\,r^2 \left[\partial_Rq_\ell(r,R)-\frac{m}{R}q_\ell(r,R)\right]_{R=a} j_\ell(k_m r_<)h_\ell^{(1)}(k_m r_>) .
  \label{eq:Fplus_uni_from_strat_new}
\end{aligned}
\end{equation}
Noting that
\begin{equation}
\int_0^\infty \dd k j_\ell(kr)j_\ell(ka) = \frac{\pi}{2\sqrt{ra}} \int_0^\infty \frac{\dd k}{k} J_\ell(kr)J_\ell(ka)
\end{equation}
and using \citep{gradshteyn2007}
\begin{equation}
\int_0^\infty \frac{dK}{K} J_\nu(Kr)J_\nu(Ka) =\frac{1}{2\nu}\left(\frac{r_<}{r_>}\right)^\nu ,
\end{equation}
we get
\begin{equation}
  q_\ell(r,a)=\frac{2(2\ell+1)}{\pi} \int_0^\infty\dd K\,j_\ell(Kr)j_\ell(Ka).
  \label{eq:q_bessel_identity_new}
\end{equation}
Additionally, comparing Eqs.~\eqref{eq:G_uniform_ansatz_new} and \eqref{eq:G_spectral_uniform_new} and using the orthogonality relation \citep[e.g.][]{Arfken1995}
\begin{equation}
\int_0^\infty r^2 \dd r j_\ell(K'r) j_\ell(Kr) = \frac{\pi}{2K^2} \delta(K-K'),
\end{equation}
yields
\begin{equation}
  \int_0^\infty\dd r\,r^2 j_\ell(Kr) \ii k_m j_\ell(k_m r_<)h_\ell^{(1)}(k_m r_>) =\frac{j_\ell(Ka)}{K^2-(k_m+\ii\epsilon)^2}.
  \label{eq:bessel_green_identity_new}
\end{equation}
and Eq.~\eqref{eq:Fplus_uni_from_strat_new} becomes
\begin{equation}
	\begin{aligned}
		\mathcal F_+^{\rm uni} &= \frac{(4\pi)^2G^2M^2\rho_0}{\sqrt2 c_s^2} \frac{2}{\pi} \sum_{\ell m}A_{\ell m}^2 \\ & \times \int_0^\infty\dd K \frac{j_\ell(Ka)}{K^2-(k_m+\ii\epsilon)^2} \left[\partial_a-\frac{m}{a}\right]j_\ell(Ka).
\end{aligned}
  \label{eq:Fplus_uni_bessel_integral_new}
\end{equation}
Using the Fourier--Bessel representation and the angular recoupling identities detailed in Appendix~\ref{app:uniform_analytic_solution}, the force reduces to
\begin{equation}
  \mathcal F_+^{\rm uni}  = \widetilde K_0 \sum_{\ell,m}\sum_{s=\pm1} \ii^s A_{\ell m}A_{\ell+s,m+1} \mathcal{H}_s(\ell,m) \mathcal R_{\ell,\ell+s}(m),
  \label{eq:app_signed_force}
\end{equation}
where
\begin{equation}
  \widetilde K_0=-\frac{(4\pi)^4\ii}{(2\pi)^3}\sqrt{\frac{4\pi}{3}}\frac{G^2M^2\rho_0}{c_s^2},
\end{equation}
\begin{equation}
  \R_{\ell_1\ell_2}(m)=\frac{\ii\pi}{2}j_{\ell_>}(m\M)\,h^{(1)}_{\ell_<}(m\M),
  \label{eq:Rell_definition}
\end{equation}
for $m>0$ and with $\ell_>=\max(\ell_1,\ell_2)$, $\ell_<=\min(\ell_1,\ell_2)$. Equation~\eqref{eq:app_signed_force} is already a closed analytic expression, but using
\begin{equation}
  A_{\ell,-m}=(-1)^mA_{\ell m}, \quad \mathcal R_{\ell,\ell+s}(-m)  =  \mathcal R_{\ell,\ell+s}(m)^*,
\end{equation}
and defining $n=m+1$, it can be rewritten as
\begin{equation}
	\begin{aligned}
  \mathcal F_+^{\rm uni} &= \widetilde{K}_{0} \sum_{\ell=0}^{\infty} \sum_{s=\pm1} \sum_{n\ge1} (-\ii s) A_{\ell,n-1}A_{\ell+s,n} \mathcal H_s(\ell,n-1) \\
			 & \times \left[\mathcal R_{\ell,\ell+s}(n)^* - \mathcal R_{\ell,\ell+s}(n-1) \right],
  \label{eq:app_final_branch_paired}
\end{aligned}
\end{equation}
which is exactly equivalent to the expression obtained by \citet{Desjacques2022} (see Sect.~\ref{sec:app_uniform}).

\section{Applications}\label{sec:applications}
\subsection{Uniform medium}\label{sec:app_uniform}

\begin{figure}
  \resizebox{\hsize}{!}{\includegraphics{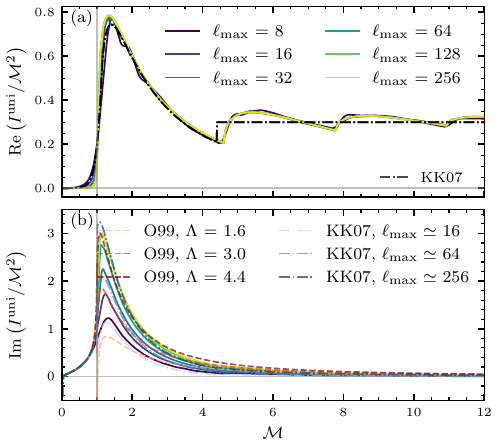}}
  \caption{$\Re\!\left(I^{\rm uni}\right)/\M^2$ (panel (a)) and $\Im\!\left(I^{\rm uni}\right)/\M^2$ (panel (b)) defined in Eq.~\eqref{eq:normalization}, as functions of the Mach number $\M$, for several truncation values of the harmonic degree $\ell_{\rm max}$. Both components agree exactly with \citet{Desjacques2022}. Dashed and dash-dotted curves show, respectively, the analytic rectilinear solution of \citet{Ostriker1999} (O99) for several values of $\Lambda=\ln(V t/b_{\rm min})$, and the fitting formula of \citet{kim2007} (KK07), for various $\ell_{\rm max} \simeq \pi a / b_{\rm min}$.}
  \label{fig:uniform_comparison}
\end{figure}

In Fig.~\ref{fig:wake_uniform}, we show the steady-state acoustic wake overdensity $\rho_1/\rho_0$ for a weak perturber orbiting in a homogeneous medium with $\M = 2$ and $\ell_{\rm max}=96$. This uniform case provides a reference solution against which the stratified calculations will be compared. To compare Eq.~\eqref{eq:app_final_branch_paired} with previous  results, we define the normalized force
\begin{equation}
\frac{I^{\rm uni}}{\M^2} = - \frac{\sqrt{2} c_s^2 \mathcal F_+^{\rm uni}}{4 \pi G^2 M^2 \rho_0},
\label{eq:normalization}
\end{equation}
and plot its real and imaginary parts as functions of the Mach number in Fig.~\ref{fig:uniform_comparison}, for several values of the truncation degree $\ell_{\rm max}$. The figure also includes the rectilinear solution of \citet{Ostriker1999} and the fitting formula of \citet{kim2007}, each shown for the corresponding values of $\Lambda=\ln(Vt/r_{\rm min})$.
For all curves shown, both $\Re\!\left(I^{\rm uni}\right)>0$ and $\Im\!\left(I^{\rm uni}\right)>0$. With our sign convention, this means that the tangential component of the gravitational force acts as a drag, while the radial component points toward the orbit centre. More importantly for the following sections, Fig.~\ref{fig:uniform_comparison} shows that the two force components have different convergence properties. The radial coefficient $\Re\!\left(I^{\rm uni}\right)$ converges rapidly with $\ell_{\rm max}$, whereas the azimuthal coefficient $\Im\!\left(I^{\rm uni}\right)$ does not converge for $\M>1$.
This non-convergence is the usual Coulomb-logarithmic sensitivity of supersonic dynamical friction to the smallest resolved scale \citep[][]{Dokuchaev1964}. For large $\ell$, the dominant contributions to $\Im\!\left(I^{\rm uni}\right)$ come from terms with $\ell\simeq n\M$, for which the spherical Bessel functions $j_\ell(n\M)$ are not exponentially suppressed. Each harmonic $n$ then contributes a term $\sim 1/\ell \simeq 1/(n\M)$, so that summing over $n=1,\ldots,\ell_{\rm max}/\M$ gives
\begin{equation}
\Im\!\left(I^{\rm uni}\right) \propto \sum_{n=1}^{\ell_{\rm max}/\M}\frac{1}{n} \sim \ln\ell_{\rm max}.
\end{equation}
This logarithmic dependence can be identified with a Coulomb logarithm $\ln(b_{\rm max}/b_{\rm min})$, where the small-scale cutoff is set by the smallest resolved scale, $b_{\rm min}=\lambda_{\rm min}\simeq \pi a/\ell_{\rm max}$, and $b_{\rm max}$ is the large-scale cutoff.
The value of $b_{\rm max}$ differs from the rectilinear case because the circular orbit imposes a discrete temporal spectrum.
Indeed, in the uniform medium considered here, the periodic source excites only the harmonics $\omega=n\Omega$, corresponding to acoustic wavenumbers $k_n=n\M/a$. There is therefore no forcing at $k<k_1=\M/a$, so the steady wake contains no power on scales larger than $b_{\rm max} = 2\pi a/\M$. This contrasts with straight-line motion, where the source excites a continuum of wavenumbers. The wake grows indefinitely and no steady state exists \citep[$b_{\rm max}=Vt$,][]{Ostriker1999}.

We verify the Coulomb-logarithmic scaling of the drag force in Fig.~\ref{fig:logI_ell}, where $\dd \Im\!\left(I^{\rm uni}\right)/\dd\ln\ell_{\rm max}\to1$ for all $\M>1$. The logarithmic dependence of $F_\varphi$ reflects the fact that each logarithmic interval between the inner and outer cutoffs contributes comparably to the drag force. In practice, the physical small-scale cutoff $b_{\rm min}$ should be set by the smallest scale on which the linear theory remains valid, namely the largest of the gravitational softening length $h_s$, the physical size of the perturber, and the scale at which the fractional density perturbation becomes $\delta=O(1)$. For $\M<1$, no such divergence occurs: $\Im\!\left(I^{\rm uni}\right)$ converges with $\ell_{\rm max}$, because a subsonic perturber does not form a Mach cone and the near-field contribution does not generate a Coulomb-logarithmic drag.

\subsection{Power-law stratification}\label{sec:power_law}
Here, we focus on the specific case
\begin{equation}
  \rho_0(r)=\rho_a\left(\frac{r}{a}\right)^p,  \quad  c_s(r)=c_a,
  \label{eq:power_law_profile}
\end{equation}
where $\rho_a\equiv \rho_0(a)$. These profiles are not intended to represent stellar envelopes, but they provide a controlled set of stratified backgrounds. The case $p=0$ recovers the homogeneous medium, while $p=-2$ corresponds to the SIS \citep{chandra39,Weinberg1986}. The latter case allows a direct comparison with \citet{Eytan2024}, who computed the acoustic wake and the associated dynamical friction in this background.
In such backgrounds, the radial acoustic Green function can be derived analytically in an analogous manner to Sect.~\ref{sec:bessel_hankel} (see appendix \ref{sec:anal_green}), yielding
\begin{equation}
  g_\ell^{(p)}(r,r';\omega)  =\frac{\ii k a^p}{\rho_a}  u_{\ell p}^{J}(r_<;k)u_{\ell p}^{H}(r_>;k),
\end{equation}
with
\begin{align}
  u_{\ell p}^{J}(r;k)
  &=\left(\frac{\pi}{2k}\right)^{1/2}r^{-\mu}J_{\nu_{\ell p}}(kr),
  u_{\ell p}^{H}(r;k)
  &=\left(\frac{\pi}{2k}\right)^{1/2}r^{-\mu}H^{(1)}_{\nu_{\ell p}}(kr),
\end{align}
and
\begin{equation}
  \mu\equiv \frac{p+1}{2},  \quad \nu_{\ell p}\equiv\left[\ell(\ell+1)+\mu^2\right]^{1/2}.
  \label{eq:power_law_indices}
\end{equation}
Setting $p=0$ results in the Bessel--Hankel Green function Eq.~\eqref{eq:G_uniform_ansatz_new}. The analytical radial Green function provides a useful benchmark for the full numerical integration of Eq.~\eqref{eq:Fplus_strat_adjacent_unpaired} and accelerates the force calculation. Again, we define the normalized force as follows
\begin{equation}
  \frac{I}{\M^2}  = -\frac{\sqrt{2}\,c_s(a)^2 \mathcal F_+}{4 \pi G^2 M^2 \rho_0(a)}.
   \label{eq:normalized_force}
\end{equation}
Throughout this subsection, we use the zero-inner-monopole convention described in Sect.~\ref{sec:DF}.

\subsubsection{Single perturber}

\begin{figure}
  \resizebox{\hsize}{!}{\includegraphics{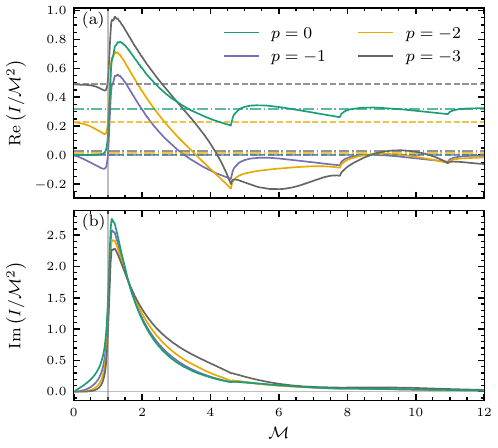}}
  \caption{$\Re\!\left(I\right)/\M^2$ (panel (a)) and $\Im\!\left(I\right)/\M^2$ (panel (b)) defined in Eq.~\eqref{eq:normalized_force}, as functions of the Mach number $\M$, for several power-law exponents $p$ of the background density profile and for harmonic degree truncation $\ell_{\rm max} = 64$, with constant background sound speed. Horizontal dashed and dash-dotted lines in panel (a) respectively correspond to the low- and high-Mach asymptotic values given by Eq.~\eqref{eq:Ir_asymp} for $\ell_{\rm max} = 64$. 
  }
  \label{fig:p_comparison}
\end{figure}

For a single perturber, we show the real and imaginary parts of $I/\M^2$ as functions of the Mach number in Fig.~\ref{fig:p_comparison}, for several values of $p$ and for $\ell_{\rm max}=64$. For $p=-2$, corresponding to the SIS, we recover the solution of \citet{Eytan2024} with the same monopole
convention; the associated wake morphology is shown in Fig.~\ref{fig:wake_powerlaw}. The low- and high-Mach limits are derived in Appendix~\ref{app:asymptotic}. We find that, in the low-Mach limit, the leading response is the static response to the instantaneous perturber position. In the high-Mach limit at fixed $\ell_{\rm max}$, we find that all modes with $m\neq0$ are suppressed. The remaining response is therefore the static response to the azimuthally averaged source, i.e. a circular ring. With the zero-inner-monopole convention, the $\ell=0$ contribution is omitted. More generally, the two limits can be written as
\begin{equation}
	{\Re\!\left[I^{\rm asy}(p;\ell_{\rm max})\right]\over \M^2} = {\chi \over p+3} +\sum_{\ell=1}^{\ell_{\max}}{\mathcal W_\ell\over 2\ell+1}
  \left[{\ell+1\over p+2\ell+3} - {\ell\over 2\ell-p-1}\right],
  \label{eq:Ir_asymp}
\end{equation}
where $\chi=0$ for the zero-inner-monopole convention and $\chi=1$ for the full static response. The angular weights are
\begin{equation}
  \mathcal W_\ell =  \begin{cases}
         1, & \M\ll1,\\
      P_\ell(0)^2, & \M\gg1\ \text{at fixed }\ell_{\max}.
  \end{cases}
\end{equation}
The corresponding leading azimuthal coefficient vanishes in both asymptotic
limits considered here,
\begin{equation}
{\Im\!\left[I^{\rm asy}(p;\ell_{\rm max})\right]\over \M^2}
  =
  \begin{cases}
      0, & \M\ll1,\\
      0, & \M\gg1\ \text{at fixed }\ell_{\max}.
  \end{cases}
\end{equation}
Resolving the physical Mach cone requires $\ell_{\max}\gtrsim\M$, so the limits $\M\to\infty$ and $\ell_{\max}\to\infty$ do not commute. We note that $p=-1$ is a special case in the low-Mach limit with the zero-inner-monopole convention. In that case, the contribution from material inside the orbit exactly cancels the contribution from material outside the orbit, mode by mode for every $\ell\ge 1$. Hence
\begin{equation}
  \lim_{\M\to 0} {\Re\!\left[I^{\rm asy}(-1;\ell_{\rm max})\right]\over \M^2} = 0.
\end{equation}
This cancellation is not a convergence property of the $\ell$-sum, but an exact cancellation within each multipole. This exact cancellation illustrates that the radial force is controlled by the relative weighting of the wake inside and outside the orbit, rather than by a local density increase near the perturber.
\begin{figure}
\resizebox{\hsize}{!}{\includegraphics{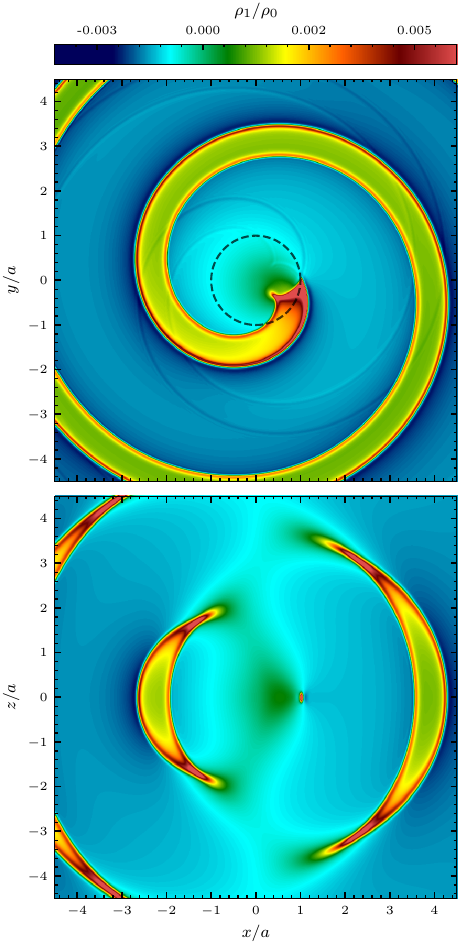}}
\caption{Steady-state acoustic wake overdensity $\rho_1/\rho_0$ on the $xy$ (top) and $xz$ (bottom) planes for a power-law exponent $p =-2$ of the background density profile with constant sound speed  (Eq.~\eqref{eq:power_law_profile}), $\M = 2$, and $\ell_{\rm max} = 96$. The perturber is located at $(x_p,y_p,z_p) = (a,0,0)$ and has a mass $M = 10^{-3}\, c_a^2 a/G$. The black circle indicates its circular orbit.}
\label{fig:wake_powerlaw}
\end{figure}

\subsubsection{Double perturbers}
\begin{figure}
  \resizebox{\hsize}{!}{\includegraphics{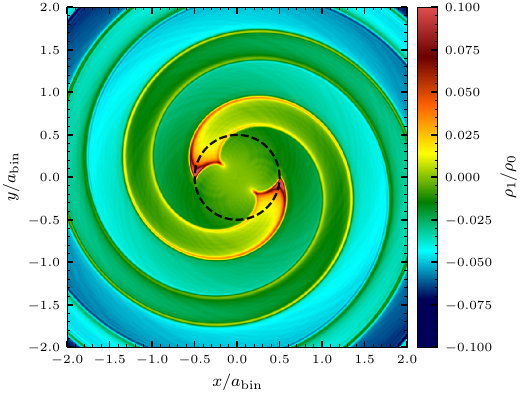}}
  \caption{Steady-state acoustic wake overdensity $\rho_1/\rho_0$ generated by a double-perturber system with $M_1 = M_2 = 10^{-3}\, c_a^2 \ab /G$ in the $xy$ plane, for a power-law background density profile with exponent $p=-2$,  orbital Mach number $\M=2$, and $\ell_{\rm max}=96$. }
  \label{fig:wake_q1_pm2}
\end{figure}

\begin{figure}
  \resizebox{\hsize}{!}{\includegraphics{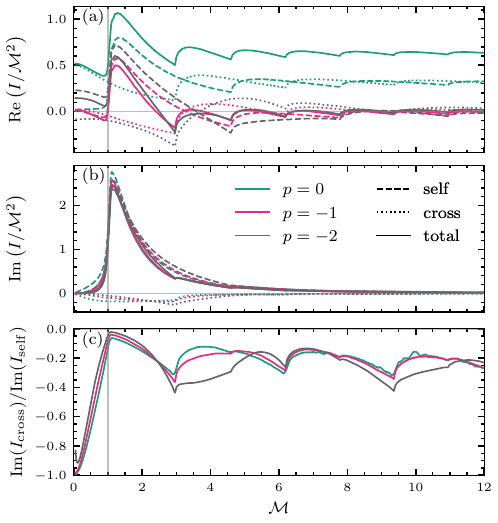}}
  \caption{$\Re\!\left(I\right)/\M^2$ (panel (a)) and $\Im\!\left(I\right)/\M^2$ (panel (b)), defined in Eq.~\eqref{eq:F1plus_binary_q}, for a double-perturber mass ratio $q=1$, as functions of the Mach number $\M$, for power-law exponents $p=0$, $-1$, and $-2$ of the background density profile and maximum harmonic degree $\ell_{\rm max}=64$, with constant background sound speed. We decompose $I$ into the contribution from the perturber's own wake, the contribution from the companion's wake, and their sum. The relative contribution to the drag force from the companion ${\rm Im}(I_{\rm cross})/{\rm Im}(I_{\rm self})$ is shown in panel (c).}
  \label{fig:double_compare_p}
\end{figure}

We now consider two perturbers on circular orbits in the same power-law backgrounds. The force on each component can be separated into the contribution from the wake raised by that component and the contribution from the wake raised by its companion, as described in Appendix~\ref{app:double}.
We illustrate the orbital-plane wake morphology for mass ratios $q=M_2/M_1=1$ and $q=0.7$ in Figs.~\ref{fig:wake_q1_pm2} and \ref{fig:wake_q07_pm2}, using $p=-2$, constant $c_s$, primary and secondary orbital Mach numbers $\M_1  = 2$ and $\M_2 = \M_1 c_s(a_1)/(q c_s(a_2))$, and $\ell_{\max}=96$.
Figure~\ref{fig:double_compare_p} shows the normalized force acting on one component of a $q=1$ double-perturber system, separated into the contribution from its own wake and that from the companion's wake.

For all $p$ values shown, we find that the total radial force exerted on the perturber is strongly modified by the companion's wake. For $p=0$, the companion's wake contribution is directed inward ($\Re(I_{\rm cross}) > 0$) over the full Mach number range shown. At low $\M$, the wake generated by the perturber is nearly symmetric and gives only a weak radial force, so the companion's wake contribution dominates. At high $\M$, the perturber's wake produces a strong inward force and the companion's wake adds a second contribution of the same sign and similar magnitude.

For $p<0$, the contribution from the companion's wake is modified by the stratified response of the medium. The density gradient changes the relative importance of the response inside and outside the orbit, and therefore changes the radial projection of the companion's wake at the perturber. The companion-wake-induced radial force becomes outward over much of the $\M$ range shown and cancels a large fraction of the inward force from the perturber's own wake for $\M \gtrsim 3$. The case $p=-1$ is particularly sensitive to the companion's wake contribution at low $\M$ because the static single-perturber radial coefficient vanishes mode by mode for all $\ell\ge1$ with the monopole convention used here.

For all $p$ shown, the companion's wake reduces the azimuthal drag. The contribution from the perturber's own wake is positive, whereas the companion's wake contribution is negative, corresponding to a forward azimuthal force, as in \citet{kim2008}. The two contributions have different cutoff dependence. The supersonic drag from the perturber's own wake contains the local Coulomb contribution from gas close to it, and therefore grows logarithmically with $\ell_{\rm max}$. The companion's wake contribution has no corresponding local part, because its source is located at a finite distance ($\ab$). It is therefore a large-scale correction to the drag, whose relative importance decreases as the local wake around the evaluated perturber is resolved to larger $\ell_{\rm max}$. The ratio $\Im(I_{\rm cross})/\Im(I_{\rm self})$ should thus be interpreted as a finite-$\ell_{\rm max}$ drag reduction, not as an asymptotic fraction independent of resolution \citep[see also][]{kim2008}.

\begin{figure}
  \resizebox{\hsize}{!}{\includegraphics{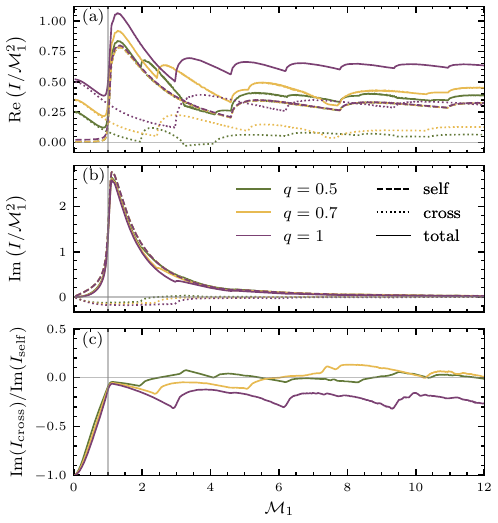}}
  \caption{Same as Fig.~\ref{fig:double_compare_p} but for fixed power-law exponent $p=0$, different mass ratios $q$,  and as a function of the more massive perturber's orbital Mach number $\M_1$.}
  \label{fig:double_compare_q}
\end{figure}

Figure~\ref{fig:double_compare_q} shows the mass-ratio dependence of the homogeneous medium double-perturber force. Decreasing $q$ weakens the companion contribution to the radial force and, near the transonic peak, reduces its effect on the azimuthal drag. The subsonic regime is different because the relevant acoustic wavelength, $\lambda_m=2\pi c_s/(m\Omega)$, is large compared with the binary separation $\ab$ for the lowest harmonics. The gas therefore responds to the large-scale binary potential rather than to two independent local wakes. The monopole is axisymmetric and gives no azimuthal force. The leading contribution to $\Im(I)$ for each component therefore comes from the dipolar part of the source potential, whose amplitude for component $i$ is proportional to $M_i a_i$\footnote{For $r\gg a$, the potential of component $i$ can be expanded as $\Phi_i(\boldsymbol{x},t) = -G M_i/|\bx-\br_i(t)| \simeq -G M_i/r - G M_i\,\br_i(t)\cdot\bx/r^3 +\mathcal{O}(G M_i a_i^2/r^3)$. The first term is independent of $\varphi$, while the dipolar term is proportional to $M_i\br_i(t)$.}. Since $M_2=qM_1$, $a_1=q\ab/(1+q)$, and $a_2=\ab/(1+q)$, one has $M_1a_1=M_2a_2$. The perturber and companion dipolar contributions therefore have equal amplitudes and opposite phases, giving $\Im(I_{\rm cross})/\Im(I_{\rm self})\simeq -1$ almost independently of $q$.

In double-perturber systems with $q < 1$, the wake of the more massive perturber can result in a positive azimuthal contribution to the force on the lighter component, $F_{\varphi,2}^{\rm cross}>0$ \citep[so called \textit{dynamical boost},][]{kim2008}, even if the lighter component's own wake gives $F_{\varphi,2}^{\rm self}<0$. The lighter body may therefore gain angular momentum from the companion's wake. This does not however imply expansion of the binary's orbit. Indeed, the separation evolution is controlled by
\begin{equation}
\dot J_{\rm orb}=a_1F_{\varphi,1}+a_2F_{\varphi,2}.
\end{equation}
For an infinite domain in the non-rotating barotropic acoustic approximation, the orbit-averaged energy balance gives
\begin{equation}
\Omega\dot J_{\rm orb}=-P_{\rm wave}\le 0,
\end{equation}
where $P_{\rm wave}$ is the outgoing acoustic luminosity. The net torque on the binary can become positive only if angular momentum is supplied to the binary, for example by the background rotation, incoming waves, or accretion.

\subsection{MESA stellar profiles}
\label{sec:mesa_profiles}

\begin{figure}
  \resizebox{\hsize}{!}{\includegraphics{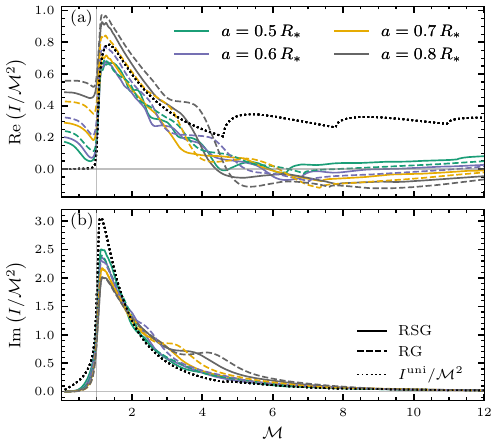}}
  \caption{$\Re\!\left(I\right)/\M^2$ (panel (a)) and $\Im\!\left(I\right)/\M^2$ (panel (b)) defined in Eq.~\eqref{eq:normalized_force}, as functions of the Mach number $\M$, for several perturber orbital radii $a$, a  maximum harmonic degree $\ell_{\rm max} = 96$, and for two giant star envelope models. The dotted lines indicate $\Re\!\left(I^{\rm uni}\right)/\M^2$ and $\Im\!\left(I^{\rm uni}\right)/\M^2$ defined in Eq.~\eqref{eq:normalization}.}
  \label{fig:mesa_I_vs_mach}
\end{figure}

We now apply the formalism to two MESA stellar-envelope profiles: a $0.77~M_\odot$ red giant (RG) at the tip of the red giant branch \citep{Kramer2020} and a $10~M_\odot$ red supergiant (RSG) \citep{Moreno2022}. The purpose is not to model a complete common-envelope event, but to evaluate how the gravitational back-reaction of the linear barotropic wake changes when the background density and adiabatic sound-speed profiles are taken from realistic giant-star envelopes. The buoyancy response associated with $N^2 \neq 0$ is not included; see Sect.~\ref{sec:buoyant_cowling_extension}.
For each model, the acoustic response and associated drag force are evaluated from Eqs.~\eqref{eq:Ylm_response_kernel}--\eqref{eq:Fplus_strat_adjacent_unpaired}, using the mass-conserving monopole constraint, Eq.~\eqref{eq:static_monopole_mass_conserving_constant}. The force is normalized as in Eq.~\eqref{eq:normalized_force}, using the local values $\rho_0(a)$ and $c_s(a)$.
Figure~\ref{fig:mesa_I_vs_mach} shows the normalized force coefficients $\Re\!\left(I\right)/\M^2$  and $\Im\!\left(I\right)/\M^2$  as functions of $\M$.
The two force components respond differently to the stellar structure. For the cases shown, the normalized azimuthal force remains close to the uniform result and depends only weakly on the stellar model and orbital radius, consistent with its Coulomb-logarithmic sensitivity to the small-scale cutoff. The radial coefficient is much more sensitive to the envelope profile: it changes with stellar model and orbital radius, and departs more strongly from the uniform-background result. This is expected if the radial force is controlled by the global imbalance between the wake inside and outside the orbit, rather than by a purely local drag contribution. The convergence tests in Figs.~\ref{fig:F_ell_RGt} and \ref{fig:F_ell_RSG} show that this behaviour is not an unresolved high-$\ell$ tail, but a property of the large-scale stratified response.

\begin{figure}
  \resizebox{\hsize}{!}{\includegraphics{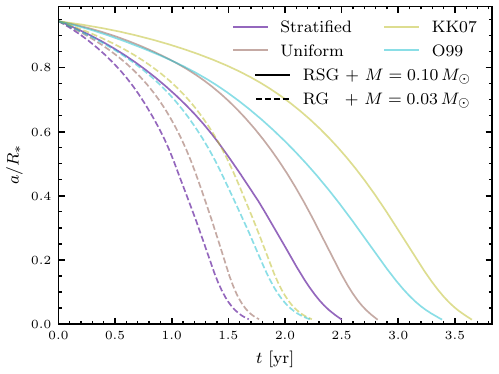
}}
\caption{Orbital decay of a low-mass perturber in the RSG (solid lines) and RG (dashed lines) MESA stellar models obtained by integrating Eq.~\eqref{eq:mesa_adot} along a sequence of circular Keplerian orbits. The stratified acoustic force is compared with different uniform-medium drag prescriptions, evaluated with the same local background quantities. The calculation assumes steady wakes, $\ell_{\max}=96$, and $\Lambda = \ln(\ell_{\rm max}/\pi)$.}
  \label{fig:a_vs_time}
\end{figure}

\begin{figure}
\resizebox{\hsize}{!}{\includegraphics{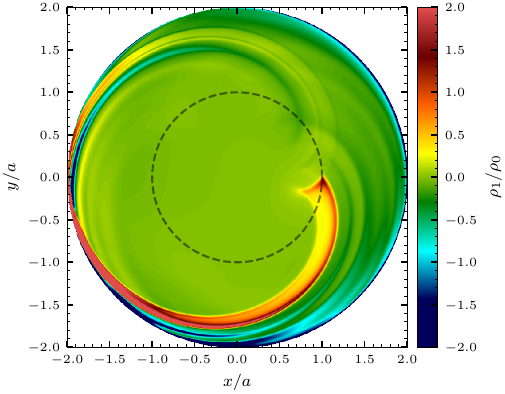}}
\caption{Same as Fig.~\ref{fig:wake_powerlaw}, but for a $M = 0.1\, M_\odot$ perturber orbiting inside the RSG stellar model at $a=0.5\,R_\ast$,  $\M_{\rm K} \simeq 1.6$ and $\ell_{\rm max} = 96$.}
\label{fig:wake_RSG}
\end{figure}

We then use the same calculation to estimate the gravitational force along a sequence of circular orbits through the envelope. At each radius $r$, we assign the Keplerian orbital angular velocity and associated Mach number, respectively
\begin{equation}
	\Omega_{\rm K}(r)   =   \left[  \frac{G M_{\rm enc}(r)}{r^3}  \right]^{1/2} \quad \text{and} \quad \M_{\rm K}(r) = {r \Omega_{\rm K}(r) \over c_s(r)},
  \label{eq:keplerian_quantities}
\end{equation}
where $M_{\rm enc}(r)$ is the enclosed mass of the MESA model at radius $r$.  Figure~\ref{fig:mesa_force_vs_a} shows the Keplerian Mach number profile as well as the force components $F_r(r)$ and $F_\varphi(r)$, for the two stellar models.
In the circular-orbit approximation, the secular orbital decay is controlled by the azimuthal component of the force. The radial component instead changes the instantaneous radial force balance and affects the orbital eccentricity \citep[][]{kim2007}. A circular orbit corrected by the radial force would satisfy
\begin{equation}
  a\Omega_{\rm eff}^2   =   \frac{G M_{\rm enc}(a)}{a^2}  -  \frac{F_r(a)}{M}.
  \label{eq:effective_orbital_frequency}
\end{equation}
Since $F_r/M \propto M$, this correction is small in the low-mass companion limit and we ignore it. The orbital angular momentum of the perturber and its change rate respectively read
\begin{equation}
	L(a)  =  M a^2\Omega_{\rm K}(a), \quad \frac{dL}{dt} =  a F_\varphi(a),
  \label{eq:orbital_angular_momentum}
\end{equation}
and the corresponding inspiral rate of the perturber is therefore
\begin{equation}
  \dot a   =   \frac{aF_\varphi(a)}{dL/da}.
  \label{eq:mesa_adot}
\end{equation}
We integrate Eq.~\eqref{eq:mesa_adot} and show the resulting wake-driven inspiral trajectories for the RSG and RG models in Fig.~\ref{fig:a_vs_time}. For the RSG model, whose steady-state acoustic wake is illustrated in Fig.~\ref{fig:wake_RSG}, we consider a perturber mass of $0.1\,M_\odot$, representative of a very low-mass main-sequence companion. For the RG model, we use $0.03\,M_\odot$, representative of a brown-dwarf companion. For comparison, we also show trajectories obtained from uniform-medium drag prescriptions, namely Eq.~\eqref{eq:app_final_branch_paired}, Kim \& Kim (\citeyear{kim2007}), and Ostriker (\citeyear{Ostriker1999}), evaluated using the same local values of $\rho_0(a)$, $c_s(a)$, and $\mathcal{M}_{\rm K}(a)$. For the latter two prescriptions, we adopt $b_{\rm min}=\pi a/\ell_{\max}$ and $b_{\rm max}=a$, matching the inner cutoff scale and using the orbital radius as the large-scale cutoff scale.
For the adopted models and cutoff, the stratified-wake calculation gives a stronger azimuthal drag than the corresponding uniform-medium prescriptions evaluated with the same local background quantities, and therefore produces a shorter acoustic-drag-driven decay time.


As shown in Sect.~\ref{sec:app_uniform}, the azimuthal force has a logarithmic sensitivity to the cutoff harmonic degree $\ell_{\max}$. This cutoff represents the smallest scale on which the linear point-mass wake is resolved. In practice, for a compact perturber, the harmonic sum should therefore be truncated once the angular scale $\pi a/\ell$ becomes smaller than either the gravitational softening scale $h_s$ or the accretion radius $R_a$ of the perturber. We define
\begin{equation}
    \begin{aligned}
        \ell_{\max}^{\rm eff}(a)
        &= { \pi a \over \max\{h_s,R_a\} } \\
        &= \min \left\{{\pi a \over h_s}, {\pi M_{\rm enc}(a) \over 2M}\left[1+\mathcal{M}_{\rm K}^{-2}(a)\right] \right\},
    \end{aligned}
    \label{eq:lmax_eff}
\end{equation}
where
\begin{equation}
    R_a = {2GM \over a^2\Omega_{\rm K}^2+c_s^2}
    \label{eq:accretion_radius}
\end{equation}
is the Bondi--Hoyle accretion radius. We adopt $h_s=3.1\,R_\odot$, corresponding to the gravitational softening length used by Kramer et al.~(\citeyear{Kramer2020}) for this red-giant model.
Below such scales the density perturbation can become nonlinear, $\rho_1/\rho_0\gtrsim 1$, and the linear acoustic calculation should no longer be trusted. Figure~\ref{fig:a_vs_time_diff_lmax} shows the acoustic wake-driven inspiral of a $0.03\,M_\odot$ perturber inside the RG envelope, for fixed truncation and for the radially varying cutoff $\ell_{\max}^{\rm eff}(a)$ from Eq.~\eqref{eq:lmax_eff}. As expected from the Coulomb-logarithm behaviour, larger fixed values of $\ell_{\max}$ result in stronger azimuthal drag and hence a faster orbital decay. The variable-cutoff decay initially follows the fixed-$\ell_{\max}$ runs with $\ell_{\max}\simeq 40$--$50$, because the accretion radius sets the smallest scale of the linear wake, rather than the softening length. As the perturber moves inward, $R_a$ decreases and eventually becomes smaller than $h_s$. From that point on, the softening length controls the cutoff and $\ell_{\max}^{\rm eff}\simeq \pi a/h_s$ decreases rapidly. The resulting trajectory therefore lies between the fixed-$\ell_{\max}$ cases at early times, but approaches the behaviour of a progressively lower-resolution wake at small radii.

For the small mass ratios considered here, tidal torques may spin up the outer envelope prior to engulfment, but full synchronization of the giant envelope is unlikely in general. The synchronization timescale scales as $q^{-2}$ in equilibrium-tide theory \citep[e.g.,][]{Hut1981}, and the angular-momentum budget can be insufficient for a Darwin-stable synchronized configuration \citep[e.g.,][]{Hut1980}. The companion should therefore enter a differentially rotating, sub-corotating envelope. We mimic this effect in a simple way, assuming that rotation only reduces the local azimuthal velocity of the gas relative to the companion. The effective Mach number is then written as
\begin{equation}
  \M_{\rm eff} = (1 - f_{\rm corot}) \M_{\rm K},
  \label{eq:effective_mach}
\end{equation}
where $f_{\rm corot} = \Omega_{\rm env} / \Omega_{\rm K}$ is the degree of corotation of the envelope with the perturber.
We show the effect of a rotating background on the orbital decay for the RG + $0.03\,M_\odot$ perturber model in Fig.~\ref{fig:adot_fcorot}. The corotation-induced orbital Mach number reduction shifts the perturber along the Mach-dependent force curve (Fig.~\ref{fig:mesa_I_vs_mach}). Since the acoustic response is strongest near the transonic regime, partial corotation can enhance the instantaneous force if it brings a supersonic perturber closer to $\M_{\rm eff}\sim 1$. Conversely, once the relative motion becomes subsonic, the drag rapidly decreases as $\M_{\rm eff}$ drops below unity.

In the model shown here, the dependence on $f_{\rm corot}$ is therefore not strictly monotonic. A small amount of corotation, $f_{\rm corot}=0.1$, slightly shortens the orbital decay time relative to the non-rotating case, because it shifts the perturber closer to the transonic peak over part of the inspiral. For larger $f_{\rm corot}$, the relative motion becomes increasingly subsonic, so the drag decreases and the inspiral slows down. The delay is small for $f_{\rm corot}=0.3$--$0.5$, but becomes dramatically longer for a nearly corotating envelope, $f_{\rm corot}=0.9$. This illustrates that the poorly constrained pre-CE tidal interaction \citep[e.g.,][]{Vick2021} can be a major uncertainty in the orbital-decay timescale.

\section{Discussion}\label{sec:discussion}
\subsection{Beyond the barotropic acoustic limit}
\label{sec:buoyant_cowling_extension}
The key assumption of the barotropic acoustic approximation is that the density perturbation is locally determined by the enthalpy perturbation,
\begin{equation}
  \rho_1=\frac{\rho_0}{c_s^2}h .
\end{equation}
This closes the problem as a single scalar equation for $h$. In a stratified background, a displaced fluid element carries its entropy through a medium whose own entropy varies with radius. The density perturbation then contains a buoyancy contribution and cannot be inferred from $h$ alone. To see this, let the background satisfy the hydrostatic equilibrium
\begin{equation}
  \nabla p_0=-\rho_0\nabla\Phi_0 ,
  \label{eq:background_hydrostatic}
\end{equation}
and introduce the Lagrangian displacement $\boldsymbol{\xi}$.  In the Cowling approximation \citep[][]{Cowling1941}, neglecting the self-gravity of the perturbation, the linearized momentum equation is
\begin{equation}
  \rho_0\frac{\partial^2\boldsymbol{\xi}}{\partial t^2}   =   -\nabla p_1-\rho_1\nabla\Phi_0-\rho_0\nabla\Phi_{\rm p}.
  \label{eq:cowling_momentum}
\end{equation}
Mass conservation gives
\begin{equation}
  \rho_1=-\nabla\cdot(\rho_0\boldsymbol{\xi}) .
\end{equation}
For adiabatic perturbations,
\begin{equation}
  \Delta p=c_{\rm ad}^2\Delta\rho,
  \quad
  c_{\rm ad}^2=\Gamma_1\frac{p_0}{\rho_0}.
  \label{eq:adiabatic_sound_speed}
\end{equation}
Writing $h=p_1/\rho_0$, this gives
\begin{equation}
  \frac{\rho_1}{\rho_0}   =   \frac{h}{c_{\rm ad}^2}   -   \boldsymbol{\xi}\cdot\boldsymbol{\mathcal A},
  \label{eq:buoyant_density_perturbation}
\end{equation}
where $\boldsymbol{\mathcal A}   \equiv   \nabla\ln\rho_0   -   \nabla\ln p_0 /\Gamma_1$.
For a spherically symmetric  background, $\boldsymbol{\mathcal A}=\mathcal A(r)\hat{\mathbf r}$, and the Brunt--Väisälä frequency
\begin{equation}
  N^2   =   g\left( \frac{1}{\Gamma_1}\frac{d\ln p_0}{dr}   -   \frac{d\ln\rho_0}{dr} \right)  = -g\mathcal A , \quad  g\equiv\frac{d\Phi_0}{dr}.
  \label{eq:brunt_vaisala}
\end{equation}
Thus $\boldsymbol{\mathcal A}=0$, equivalently $N^2=0$, is the neutrally stratified limit. After expanding in spherical harmonics and Fourier modes, the buoyant Cowling problem is therefore not a scalar equation for $h_{\ell m}$.  It is a coupled radial problem for the enthalpy perturbation and the radial displacement.  With
\begin{equation}
  \boldsymbol{\xi}   =   \xi_{r,\ell m}(r,\omega)Y_\ell^m\hat{\mathbf r}   +   \xi_{\perp,\ell m}(r,\omega)\nabla_\Omega Y_\ell^m ,
  \label{eq:displacement_decomposition}
\end{equation}
the horizontal momentum equation gives
\begin{equation}
  \xi_{\perp,\ell m}   =   \frac{h_{\ell m}+\Phi_{{\rm p},\ell m}}{\omega^2 r}, 
\end{equation}
while the radial momentum and continuity equations give
\begin{equation}
  \left(\omega^2-N^2\right)\xi_{r,\ell m}   =   \frac{d}{dr}\left(h_{\ell m}+\Phi_{{\rm p},\ell m}\right)   +   \mathcal A h_{\ell m},
  \label{eq:buoy_radial_momentum_compact}
\end{equation}
and
\begin{equation}
  \frac{1}{r^2}\frac{d}{dr}   \left(r^2\rho_0\xi_{r,\ell m}\right) = \rho_0\left[ -\frac{h_{\ell m}}{c_{\rm ad}^2} +\mathcal A\xi_{r,\ell m} +\frac{\ell(\ell+1)}{\omega^2r^2} \left(h_{\ell m}+\Phi_{{\rm p},\ell m}\right)\right].
  \label{eq:buoy_continuity_compact}
\end{equation}
The barotropic acoustic formalism is recovered by setting
\begin{equation}
  \mathcal A=0, \quad N^2=0, \quad c_{\rm ad}^2=c_s^2 .
\end{equation}
Equations~\eqref{eq:buoy_radial_momentum_compact} and \eqref{eq:buoy_continuity_compact} are the standard displacement-based Cowling equations used to describe tidally forced internal gravity waves in stratified stellar \citep[e.g.,][]{Press1981,Ahuir2021} and planetary \citep[e.g.,][]{Dhouib2024} interiors. Closely related formulations are also used to study tides in stratified planetary atmospheres \citep[e.g.,][]{Auclair2017}.

\subsection{Acoustic propagation in stellar envelopes}
We first consider where the computed acoustic response can actually propagate. For a spherical harmonic of degree $\ell$, the Lamb frequency measuring the acoustic frequency associated with the horizontal wavelength of the mode reads
\begin{equation}
  S_\ell(r)=\frac{\sqrt{\ell(\ell+1)}c_s(r)}{r}.
  \label{eq:lamb_frequency}
\end{equation}
A pressure wave can propagate radially only when its forcing frequency is sufficiently large compared with $S_\ell$, otherwise the radial wavenumber becomes imaginary and the response is evanescent \citep[e.g.,][]{Maeder2009}. At the orbital radius,
\begin{equation}
  \frac{m\Omega}{S_\ell(a)}
  =
  \frac{m\mathcal{M}}{\sqrt{\ell(\ell+1)}} .
\end{equation}
Sectoral harmonics ($m=\ell$) are therefore the most favourable to acoustic propagation, with $m\Omega/S_\ell(a)\simeq \mathcal{M}$ for large $\ell$.

\begin{figure}
  \resizebox{\hsize}{!}{\includegraphics{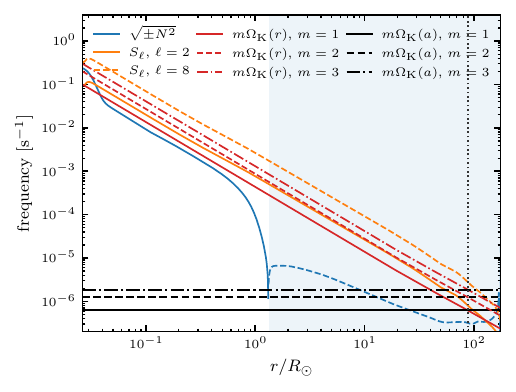}}
  \caption{Propagation diagram showing the Brunt--Väisälä frequency $\sqrt{|N^2|}$, the Lamb frequencies $S_\ell$ for $\ell=2$ and $8$, and the forcing frequencies $m\Omega$ for $m=1,2,3$, for the $0.03\,M_\odot$ perturber located at $a=0.5\,R_\ast$ in the RG MESA stellar model. Dashed blue segments correspond to $N^2<0$, and the dotted vertical line marks the perturber orbit.}
  \label{fig:mesa_frequencies_RGt}
\end{figure}

Figures~\ref{fig:mesa_frequencies_RGt} and \ref{fig:mesa_frequencies_RSG} show the radial profiles of the Brunt--Väisälä frequency $N$ and the Lamb frequencies $S_\ell$ for the RG and RSG models, respectively. Two sets of orbital frequencies are included. The horizontal lines show the forcing frequencies $m\Omega_{\rm K}(a)$ produced by a perturber fixed at $a=0.5R_\ast$. The radially varying curves show  the forcing frequencies $m\Omega_{\rm K}(r)$ that would be obtained for a sequence of circular orbits with $a=r$.  For the fixed orbit, three regions can be distinguished. In the inner radiative region, the frequencies $m\Omega_{\rm K}(a)$ shown lie below both $N$ and $S_\ell$. The acoustic response is therefore radially evanescent there and the density perturbation produced by entropy advection cannot be neglected. The companion can then excite a buoyancy-driven response, including internal gravity waves, which transport angular momentum through the radiative layer and exert a torque that is not accounted for by the present barotropic acoustic model (see Sect.~\ref{sec:buoyant_cowling_extension}).
The curves $m\Omega_{\rm K}(r)$ are very close to $S_\ell(r)$ and slightly above $N(r)$ throughout the radiative region. This indicates that a perturber orbiting there would lie near the transition between two propagation regimes. Once its orbital radius $a$ is fixed, its forcing frequency $m\Omega_{\rm K}(a)$ is constant with radius, whereas $N(r)$ and $S_\ell(r)$ increase inward and decrease outward. The forcing frequency would therefore tend to lie below $N$ and $S_\ell$ inside the orbit and above them outside the orbit, favouring a buoyancy-driven response at $r<a$ and acoustic propagation at $r>a$. 
In the extended convective envelope, $N^2\lesssim0$, so there is no stable gravity-wave cavity and the relevant acoustic propagation criterion is the Lamb condition, $m\Omega_{\rm K}(a)>S_\ell(r)$. For sectoral harmonics, this condition is satisfied locally near the orbit when $\M \gtrsim1$, whereas lower-$m$ components can remain Lamb-evanescent. Farther outward, the decrease of $S_\ell$ allows an increasing number of modes to propagate as acoustic waves.

\subsection{Limits of linear response}

In this work, we found that, in the supersonic regime, the azimuthal force is sensitive to the smallest scale on which the perturbed flow is resolved, whereas the radial force is controlled by the large-scale asymmetry of the wake.
This makes the azimuthal force more difficult to compare with hydrodynamical simulations than the radial force. Hydrodynamical simulations do not resolve a true point-mass wake down to arbitrarily small scales: the effective inner cutoff is set by the softening length, the grid scale, and by the radius inside which the response becomes nonlinear. The linear result should therefore be compared to simulations only after choosing an $\ell_{\max}$ that represents this effective cutoff (see Sect.~\ref{sec:mesa_profiles}).

The linear approximation fails where $|\rho_1|/\rho_0\gtrsim1$.
One expected location of linear approximation breakdown is the gas close to the perturber, typically on scales smaller than $b_{\rm min}$, the largest of the accretion radius, the physical radius of the perturber, the softening length. For the linear wake to be cleanly separated from this unresolved near-field region, $b_{\rm min}$ should be small compared with the characteristic scales of the problem: the orbital radius $a$ and the background density scale height $H_\rho$. Thus we require
\begin{equation}
b_{\rm min } \ll a,\quad b_{\rm min} \ll H_\rho .
\end{equation}
When these conditions are satisfied, the nonlinear flow occupies only a small region around the companion, while the large-scale wake can still be treated linearly. 
Low-mass compact companions naturally favour $b_{\rm min}\ll a$. However, $b_{\rm min}\ll H_\rho$ may fail in the outermost layers of giant-star envelopes, where $H_\rho$ becomes small.
In the absence of accretion, gas inside this region may form a nearly hydrostatic envelope around the companion \citep[e.g.,][]{kim2009,Thun2016,Prust2024,Gagnier2025,Gagnier2026} that contributes little to the net drag. Instead, it defines an effective inner cutoff for the resolved wake. For planet perturbers, the accretion radius can be smaller than their physical radius. In that case ram-pressure drag on the planetary surface may dominate over gravitational dynamical friction \citep[e.g.,][]{Lau2025}, and ablation can become central to the evolution \citep[e.g.,][]{Jia2018,Lau2026}.
Nonlinear response is not restricted to the near field. In a stratified envelope, acoustic waves propagating into regions of decreasing background density can reach large fractional amplitudes, $|\rho_1|/\rho_0 \gtrsim 1$ (Fig.~\ref{fig:wake_RSG}). In these regions, the wave may steepen into a shock and the corresponding solution can no longer be interpreted as a linear response.

\subsection{Steady circular-orbit approximation}
The inspiral calculation presented in Sect.~\ref{sec:mesa_profiles} treats the orbital evolution as a sequence of steady states. This quasi-steady approximation requires the acoustic response to adjust on a timescale short compared with the orbital-evolution timescale and does not capture the time-dependent coupling between the orbit and the wake. A natural next step is a self-consistent time-dependent calculation that evolves the orbit together with the retarded wake.
\section{Conclusion}
\label{sec:conclu}

We have formulated the linear barotropic acoustic response of a point perturber on a circular orbit in a hydrostatic, spherically stratified gaseous medium. Expanding the response in spherical harmonics and orbital Fourier modes, we expressed the gravitational back-reaction of the wake as an adjacent-multipole sum whose radial response coefficients are computed from retarded acoustic Green functions. The formulation recovers the homogeneous circular-orbit problem as a limiting case and provides a flexible tool for computing the gravitational back-reaction of the linear barotropic wake for arbitrary prescribed radial profiles of density and sound speed.

We found that stratification modifies dynamical friction by changing the global acoustic wake, not only by changing the local density, sound speed, and Mach number at the perturber position. The radial component of the force is set by the low-order, orbit-scale asymmetry of the wake, whereas the azimuthal component retains the Coulomb-logarithmic drag of the homogeneous problem in the supersonic regime and remains sensitive to the small-scale cutoff. The gravitational back-reaction of the linear barotropic wake in a stratified envelope therefore combines an envelope-scale response with a cutoff-sensitive contribution that must be fixed by a physically-motivated cutoff scale.

For power-law density profiles with constant sound speed, the response admits an analytic Green-function solution and recovers the SIS result of \citet{Eytan2024}. The same formalism also applies to double perturbers, where the companion wake can substantially modify the radial force and reduce the azimuthal drag on a given component, without introducing an additional local Coulomb-logarithmic term.  In the MESA giant-envelope models, the radial force depends strongly on the orbital radius and the stellar structure, whereas the azimuthal drag remains closer to the homogeneous medium result. In the examples considered in this work, the stratified acoustic drag results in shorter inspiral times than uniform-medium prescriptions evaluated with the same local background quantities.

The linear acoustic contribution to dynamical friction inside stellar envelopes is therefore not, in general, a purely local process. Even within the linear acoustic regime, the global structure of the envelope can modify the wake and its back-reaction on the perturber. Computing this response for embedded perturbers is a step toward computationally efficient models of common-envelope and planetary-engulfment inspirals, but several ingredients remain to be added.
This should be complemented by local three-dimensional hydrodynamical simulations to quantify the nonlinear gravitational response near the perturber and isolate the near-field contribution not captured by the linear theory.
The formalism presented in this work is versatile and could be adapted to other embedded-perturber problems, such as planet and planetesimal migration in gaseous disks.

\begin{acknowledgements}
D.G. thanks R. Andrassy for helpful comments on the manuscript and  acknowledges support by the Klaus-Tschira Foundation and funding by the European Union (ERC, ExCEED, project number 101096243). Views and opinions expressed are, however, those of the author only and do not necessarily reflect those of the European Union or the European Research Council Executive Agency. Neither the European Union nor the granting authority can be held responsible for them. This work has received funding from the European Research Council (ERC) under the European Union's Horizon 2020 research and innovation program (Grant agreement No. 945806). Software: NumPy \citep{Numpy}, SciPy \citep{Scipy}, Matplotlib \citep{Matplotlib}.
\end{acknowledgements}

\section*{Data availability}
Tools to reproduce the findings of this study are openly available at: \url{https://github.com/dgagnier/DF_stratified}. 

\bibliographystyle{aa}
\bibliography{bibnew}

\begin{appendix}
\nolinenumbers
\clearpage
\begin{table*}[!t]
\section{Summary of notations}
\label{app:notation}
\begingroup
\renewcommand{\arraystretch}{1.02}
\caption{Summary of the notations used in this work.}
\label{tab:notation}
\begin{tabularx}{\textwidth}{@{}>{\raggedright\arraybackslash}p{0.18\textwidth}>{\raggedright\arraybackslash}X>{\raggedright\arraybackslash}p{0.13\textwidth}@{}}
\hline\hline
Symbol & Definition & Introduced in \\
\hline

\multicolumn{3}{@{}l}{\textit{Physical setup}}\\

$G$
& Gravitational constant
& Eq.~\eqref{eq:force_general} \\

$M$
& Mass of the perturber
& Sect.~\ref{sec:setup} \\

$a$
& Radius of the perturber's circular orbit
& Eq.~\eqref{eq:orbit_position} \\

$\boldsymbol{r}_{\rm p}(t)$
& Instantaneous position of the perturber
& Eq.~\eqref{eq:orbit_position} \\

$\Omega$
& Orbital angular velocity
& Eq.~\eqref{eq:orbit_position} \\

$\mathcal{M}$
& Orbital Mach number, $\mathcal{M}=\Omega a/c_s(a)$
& Eq.~\eqref{eq:orbital_mach} \\

$\rho_0(r),\,p_0(r),\,c_s(r)$
& Background density, pressure, and sound speed
& Sect.~\ref{sec:setup} \\

$\rho_1,\,p_1$
& Wake density and pressure perturbations
& Sect.~\ref{sec:setup} \\

$\boldsymbol{v}$
& Eulerian velocity perturbation
& Eq.~\eqref{eq:strat_cont_h} \\

$\Phi_{\rm p}$
& Gravitational potential of the perturber
& Eq.~\eqref{eq:strat_mom} \\

$h$
& Enthalpy perturbation, $h=p_1/\rho_0$
& Eq.~\eqref{eq:h_def} \\

$\boldsymbol{F}$
& Gravitational force exerted by the wake on the perturber
& Eq.~\eqref{eq:force_general} \\

$F_r,\,F_\varphi,\,F_z$
& Radial, azimuthal, and vertical components of $\boldsymbol{F}$
& Sect.~\ref{sec:setup} \\

$\hat{\boldsymbol e}_{+},\,\mathcal F_{+}$
& Positive-helicity basis vector and complex force amplitude
& Eq.~\eqref{eq:Fplus_def} \\

\multicolumn{3}{@{}l}{\textit{Harmonic expansion and acoustic response}}\\

$\ell,\,m$
& Spherical-harmonic degree and azimuthal order
& Eq.~\eqref{eq:harmonic_expansion} \\

$Y_\ell^m$
& Spherical harmonic in the Condon--Shortley convention
& Eq.~\eqref{eq:harmonic_expansion} \\

$\omega$
& Fourier angular frequency
& Eq.~\eqref{eq:fourier_decomposition} \\

$h_{\ell m}(r;\omega)$
& Radial Fourier--harmonic coefficient of $h$
& Eq.~\eqref{eq:fourier_decomposition} \\

$\Phi_{\ell m}(r;\omega)$
& Radial Fourier--harmonic coefficient of $\Phi_{\rm p}$
& Eq.~\eqref{eq:fourier_decomposition} \\

$\mathcal{L}_{\ell}$
& Radial acoustic operator
& Eq.~\eqref{eq:Lell_def} \\

$g_{\ell}(r,r';\omega)$
& Retarded radial acoustic Green function
& Eq.~\eqref{eq:Green_strat} \\

$\Psi$
& Gravitational potential generated by the wake
& Eq.~\eqref{eq:wake_potential_def} \\

$q_\ell(r,R)$
& Radial kernel in the Laplace expansion,
$q_\ell=r_<^\ell/r_>^{\ell+1}$
& Eq.~\eqref{eq:laplace_expansion} \\

$r_<,\,r_>$
& Smaller and larger radial arguments entering $q_\ell$
& Eq.~\eqref{eq:laplace_expansion} \\

$\psi_{\ell m}(R;\omega)$
& Radial Fourier amplitude of the wake gravitational potential
& Eqs.~\eqref{eq:wake_potential_time_dependence}--\eqref{eq:wake_potential_amplitude} \\

$H_{\ell m}(r;\omega)$
& Radial Green-function response to the source $\Phi_{\ell m}^{(0)}(r)$
& Eq.~\eqref{eq:Hlm_def_general} \\

$\Phi_{\ell m}^{(0)}(r)$
& Same radial coefficient as $\Phi_{\ell m}(r,t)$, but without the factor $e^{-im\Omega t}$
& Eq.~\eqref{eq:Phi0_q_def} \\

$A_{\ell m}$
& Orbital-plane spherical-harmonic coefficient,
$Y_\ell^{m*}(\pi/2,0)$
& Eq.~\eqref{eq:Alm_def} \\

$\mathfrak{F}_{\ell m}$
& Radial force integral entering the harmonic expansion of $F_+$
& Eq.~\eqref{eq:Frak_lm_def} \\

$\mathcal{J}_{\ell,+},\,\mathcal{J}_{\ell,-}$
& Contributions to $\mathfrak{F}_{\ell m}$ from radii inside and outside
the perturber orbit
& Eqs.~\eqref{eq:Jell_plus}--\eqref{eq:Jell_minus} \\

$\mathcal{Y}_{\ell m}(r;\omega)$
& Green-function response kernel entering the force integrals
& Eq.~\eqref{eq:Ylm_response_kernel} \\

$w(r)$
& Radial weight, $w=\rho_0/c_s^2$
& Eq.~\eqref{eq:Ylm_response_kernel} \\

$D_\ell$
& Angular normalization coefficient in the adjacent-multipole force
expansion
& Eq.~\eqref{eq:Dell_def} \\

$\mathcal{H}_{s}(\ell,m)$
& Helicity recoupling coefficient coupling the $\ell$ and $\ell+s$
multipoles, with $s=\pm1$
& Eq.~\eqref{eq:Hs_single_closed} \\

$C$
& Constant fixing the global monopole convention
& Eqs.~\eqref{eq:static_monopole_general}--\eqref{eq:static_monopole_mass_conserving_constant} \\

\multicolumn{3}{@{}l}{\textit{Uniform and power-law backgrounds}}\\

$\mathcal{D}_\ell$
& Spherical radial Laplacian
& Eq.~\eqref{eq:Dell_uniform} \\

$k$
& Acoustic radial wavenumber, $k=\omega/c_s$
& Eq.~\eqref{eq:G_uniform_equation_D_new} \\

$k_m$
& Wavenumber of the $m$th orbital harmonic,
$k_m=m\Omega/c_s$
& Eq.~\eqref{eq:G_uniform_ansatz_new} \\

$j_\ell,\,h_\ell^{(1)}$
& Spherical Bessel function and outgoing spherical Hankel function
& Eq.~\eqref{eq:G_uniform_ansatz_new} \\

$\mathcal{R}_{\ell_1\ell_2}(m)$
& Uniform-medium radial integral entering the analytic force
& Eq.~\eqref{eq:Rell_definition} \\

$I^{\rm uni}$
& Dimensionless dynamical-friction force coefficient in a uniform medium
& Eq.~\eqref{eq:normalization} \\

$p$
& Power-law exponent of the background density,
$\rho_0(r)\propto r^p$
& Eq.~\eqref{eq:power_law_profile} \\

$\rho_a,\,c_a$
& Density and sound speed evaluated at the orbital radius $a$
& Eq.~\eqref{eq:power_law_profile} \\

$\mu,\,\nu_{\ell p}$
& Indices used in the analytic power-law Green function
& Eq.~\eqref{eq:power_law_indices} \\

$I$
& Dimensionless dynamical-friction force coefficient in a stratified
background
& Eq.~\eqref{eq:normalized_force} \\

\multicolumn{3}{@{}l}{\textit{Stellar-envelope and inspiral quantities}}\\
$M_{\rm enc}(r)$
& Stellar mass enclosed within radius $r$
& Eq.~\eqref{eq:keplerian_quantities} \\

$\Omega_{\rm K},\,\mathcal{M}_{\rm K}$
& Keplerian orbital angular velocity and corresponding Mach number
& Eq.~\eqref{eq:keplerian_quantities} \\

$\Omega_{\rm eff}$
& Circular-orbit angular velocity including the radial wake force
& Eq.~\eqref{eq:effective_orbital_frequency} \\

$L$
& Orbital angular momentum of the perturber
& Eq.~\eqref{eq:orbital_angular_momentum} \\

$\dot a$
& Secular inspiral rate
& Eq.~\eqref{eq:mesa_adot} \\

$h_s$
& Gravitational softening scale used to set the small-scale cutoff
& Eq.~\eqref{eq:lmax_eff} \\

$R_a$
& Bondi--Hoyle accretion radius
& Eq.~\eqref{eq:accretion_radius} \\

$\ell_{\max}$
& Maximum spherical-harmonic degree retained in the response
& Sect.~\ref{sec:applications} \\

$\ell_{\max}^{\rm eff}$
& Effective harmonic cutoff associated with $h_s$ and $R_a$
& Eq.~\eqref{eq:lmax_eff} \\

\hline
\end{tabularx}
\endgroup
\end{table*}

\twocolumn[{
\begingroup
\renewcommand{\arraystretch}{1.02}
\small\textbf{Table~\ref{tab:notation}.} Summary of the notations used in this work (continued).\\

\begin{tabularx}{\textwidth}{@{}>{\raggedright\arraybackslash}p{0.18\textwidth}>{\raggedright\arraybackslash}X>{\raggedright\arraybackslash}p{0.13\textwidth}@{}}
\hline\hline
Symbol & Definition & Introduced in \\
\hline

$\mathcal{M}_{\rm eff}$
& Effective Mach number in a partially corotating envelope
& Eq.~\eqref{eq:effective_mach} \\

$f_{\rm corot}$
& Degree of envelope corotation,
$f_{\rm corot}=\Omega_{\rm env}/\Omega_{\rm K}$
& Eq.~\eqref{eq:effective_mach} \\
\multicolumn{3}{@{}l}{\textit{Non-barotropic extension}}\\

$\Phi_0$
& Background gravitational potential
& Eq.~\eqref{eq:background_hydrostatic} \\

$\boldsymbol{\xi}$
& Lagrangian fluid displacement
& Eq.~\eqref{eq:cowling_momentum} \\

$c_{\rm ad}$
& Adiabatic sound speed,
$c_{\rm ad}^2=\Gamma_1 p_0/\rho_0$
& Eq.~\eqref{eq:adiabatic_sound_speed} \\

$\Gamma_1$
& First adiabatic exponent
& Eq.~\eqref{eq:adiabatic_sound_speed} \\

$\boldsymbol{\mathcal A}$
& Schwarzschild-discriminant vector,
$\boldsymbol{\mathcal A}=\nabla\ln\rho_0-\nabla\ln p_0/\Gamma_1$
& Eq.~\eqref{eq:buoyant_density_perturbation} \\

$N$
& Brunt--V\"ais\"al\"a frequency
& Eq.~\eqref{eq:brunt_vaisala} \\

$\xi_{r,\ell m},\,\xi_{\perp,\ell m}$
& Radial and horizontal harmonic components of the Lagrangian displacement
& Eq.~\eqref{eq:displacement_decomposition} \\

$S_\ell$
& Lamb frequency
& Eq.~\eqref{eq:lamb_frequency} \\

\hline
\end{tabularx}
\endgroup
\vspace{1em}
}]

\section{Numerical evaluation of the radial response}
\label{app:numerical_implementation}
For constant sound speed power-law models presented in Sect.~\ref{sec:power_law}, we evaluate the Green-function integral \eqref{eq:Ylm_response_kernel}  as
\begin{align}
\mathcal{Y}_{\ell m}(r;\omega) &= C_{\ell p}(k)\Bigg[ u^H_{\ell p}(r;k) \int_{r_{\rm min}}^r r'^2 u^J_{\ell p}(r';k)\,\mathcal{L}_\ell q_\ell(r',a)\,\mathrm{d}r'
  \notag\\
  &+   u^J_{\ell p}(r;k) \int_r^{r_{\rm max}}  r'^2 u^H_{\ell p}(r';k)\,\mathcal{L}_\ell q_\ell(r',a)\,\mathrm{d}r'
  \Bigg],
\end{align}
on a logarithmic grid. Since $q_\ell$ is continuous at $r=a$ but its derivative is not (see Sect.~\ref{sec:enthalpy_response}), the source contains a smooth stratification-dependent term and a point-source term,
\begin{equation}
  \mathcal{L}_\ell  q_\ell = -\rho_0'(r) \partial_r q_\ell + \rho_0(a)\frac{2\ell+1}{a^2}\delta(r-a).
\end{equation}
The smooth term is integrated on the radial grid.  The delta term is inserted exactly in the power-law Green-function calculation, using the identity $\int f(r')\delta(r'-a) \dd r'=f(a)$. 
In the case of arbitrary background profiles calculation (e.g., Sect.~\ref{sec:mesa_profiles}) there is no closed expression for $g_\ell(r,r';\omega)$.  We therefore compute $\mathcal{Y}_{\ell m}$ by solving the radial equation on the grid,
\begin{equation}
  \left[\mathcal{L}_\ell-(m\Omega+\ii\epsilon)^2 w(r) \right] \mathcal{Y}_{\ell m}   =  \mathcal{L}_\ell q_\ell.
\end{equation}
The replacement $\omega \rightarrow \omega+\ii\epsilon$ selects the outgoing Green function and (weakly) damps the formally infinite oscillatory tail ($h_\ell^{(1)}(kr) \sim e^{+\ii kr}/r$ at large $r$) on the finite numerical grid.  We typically choose $\epsilon = 10^{-3}\, c_s(a)/a$.

In the power-law case, the Green function expression already contains the boundary conditions. In the case of arbitrary background  profiles calculation, we explicitly enforce regularity toward the origin, and no incoming wave from infinity.  At the inner boundary, two possible multipole behaviours are possible
\begin{equation}
  \mathcal{Y}_{\ell m} \sim A r^\ell + B r^{-\ell-1} .
\end{equation}
The second term diverges at the origin. We thus set $B=0$ by imposing 
\begin{equation}
	\partial_r\mathcal{Y}_{\ell m} = \frac{\ell}{r}\mathcal{Y}_{\ell m}, \quad \text{at } r= r_{\rm min} .
\end{equation}
At the outer boundary, for $m\ne0$, we force the solution to have the slope of an outgoing wave $ \mathcal{Y}_{\ell m}(r_{\rm max};\omega) \propto h_\ell^{(1)}(kr_{\rm max}) \sim e^{+ikr_{\rm max}}/r_{\rm max}$, that is 
\begin{equation}
\partial_r\mathcal{Y}_{\ell m} = k \frac{h_\ell^{(1)\prime}(kr)}{h_\ell^{(1)}(kr)} \mathcal{Y}_{\ell m}, \quad \text{at } r= r_{\rm max} ,
\end{equation}
where $k \equiv \omega / c_s(r_{\rm max})$ and the prime symbol denotes derivative  with respect to the argument $kr$.
The $m=0$ case is treated separately using the static solution and the chosen monopole convention.

\section{Analytic reduction of the force for a uniform medium}
\label{app:uniform_analytic_solution}

Equation~\eqref{eq:Fplus_uni_bessel_integral_new} gives the uniform-medium force, but it is not yet the final analytic solution because the helicity derivative still acts on the Bessel function. In this appendix we reduce Eq.~\eqref{eq:Fplus_uni_bessel_integral_new} to the adjacent-multipole form Eq.~\eqref{eq:app_signed_force}.

\subsection{Recoupling the helicity derivative}

We first derive the angular identity that replaces the differentiated  Bessel function by adjacent multipoles.  On the orbital plane,
\begin{equation}
  \hat{\bm e}_+\cdot\nabla_{\bm R} = \frac{\ee^{\ii\varphi}}{\sqrt2} \left( \partial_R+\frac{\ii}{R}\partial_\varphi \right).
\end{equation}
Therefore, at $\bm R=a\hat{\bm x}$,
\begin{equation}
  \left. \hat{\bm e}_+\cdot\nabla_{\bm R}  \left[ j_\ell(KR)Y_\ell^m(\hat{\bm R}) \right] \right|_{\bm R=a\hat{\bm x}} = \frac{A_{\ell m}}{\sqrt2}  \left( \partial_a-\frac{m}{a} \right)j_\ell(Ka),
  \label{eq:app_direct_derivative}
\end{equation}
where $A_{\ell m}=Y_\ell^{m*}(\pi/2,0)$.  Now, using the Rayleigh expansion
\begin{equation}
  \ee^{\ii\bm K\cdot\bm R} = 4\pi \sum_{\ell,m} \ii^\ell j_\ell(KR) Y_\ell^{m *}(\hat{\bm K})Y_\ell^m(\hat{\bm R}),
  \label{eq:app_rayleigh}
\end{equation}
applying $\hat{\bm e}_+\cdot\nabla_{\bm R}$ term by term and then using Eq.~\eqref{eq:app_direct_derivative} gives
\begin{equation}
  \left.  \hat{\bm e}_+\cdot\nabla_{\bm R} \ee^{\ii\bm K\cdot\bm R} \right|_{\bm R=a\hat{\bm x}} = 4\pi\sum_{\ell,m} \ii^\ell Y_\ell^{m *}(\hat{\bm K}) \frac{A_{\ell m}}{\sqrt2} \left(\partial_a-\frac{m}{a}\right)j_\ell(Ka).
  \label{eq:app_derivative_first_way}
\end{equation}
The same quantity can also be obtained by differentiating the plane wave first:
\begin{equation}
  \hat{\bm e}_+\cdot\nabla_{\bm R} \ee^{\ii\bm K\cdot\bm R}= \ii (\hat{\bm e}_+\cdot\bm K)\ee^{\ii\bm K\cdot\bm R}.
\end{equation}
With our Condon--Shortley convention,
\begin{equation}
  Y_1^1(\hat{\bm K}) = -\sqrt{\frac{3}{8\pi}}  \sin\theta_K\,\ee^{\ii\varphi_K},
\end{equation}
and therefore
\begin{equation}
  \hat{\bm e}_+\cdot\bm K = \frac{K}{\sqrt2}\sin\theta_K \ee^{\ii\varphi_K}  = -K \sqrt{\frac{4\pi}{3}}Y_1^1(\hat{\bm K}).
\end{equation}
Thus
\begin{equation}
  \hat{\bm e}_+\cdot\nabla_{\bm R} \ee^{\ii\bm K\cdot\bm R}  =-\ii K\sqrt{\frac{4\pi}{3}} Y_1^1(\hat{\bm K}) \ee^{\ii\bm K\cdot\bm R}.
\end{equation}
Expanding the remaining plane wave with Eq.~\eqref{eq:app_rayleigh} and then setting $\bm R=a\hat{\bm x}$ gives
\begin{equation}
  \left.  \hat{\bm e}_+\cdot\nabla_{\bm R} \ee^{\ii\bm K\cdot\bm R} \right|_{\bm R=a\hat{\bm x}}= -4\pi\ii K\sqrt{\frac{4\pi}{3}} \sum_{\lambda,\mu} \ii^\lambda A_{\lambda\mu} j_\lambda(Ka)  Y_1^1(\hat{\bm K})  Y_\lambda^{\mu *}(\hat{\bm K}).
  \label{eq:app_derivative_second_way}
\end{equation}
Equations~\eqref{eq:app_derivative_first_way} and \eqref{eq:app_derivative_second_way} are two expansions of the same function of $\hat{\bm K}$. We project both sides onto $Y_\ell^{m*}(\hat{\bm K})$ by multiplying by $Y_\ell^m(\hat{\bm K})$ and integrating over $\dd^2\hat{\bm K}$. The first expansion gives, by orthogonality,
\begin{equation}
  4\pi \ii^\ell \frac{A_{\ell m}}{\sqrt2} \left(\partial_a-\frac{m}{a}\right)j_\ell(Ka).
\end{equation}
The second expansion gives
\begin{equation}
  -4\pi\ii K\sqrt{\frac{4\pi}{3}} \sum_{\lambda,\mu} \ii^\lambda A_{\lambda\mu} j_\lambda(Ka) \int \dd^2\hat{\bm K} Y_\ell^m Y_1^1 Y_\lambda^{\mu *}.
\end{equation}
Using $Y_\lambda^{\mu *}=(-1)^\mu Y_\lambda^{-\mu}$, the angular integral becomes a Gaunt integral. Its azimuthal selection rule gives $\mu=m+1$. The triangle rule allows $\lambda=\ell-1,\ell,\ell+1$, but parity eliminates $\lambda=\ell$. Hence
\begin{equation}
  \lambda=\ell+s,\quad \mu=m+1,\quad s=\pm1.
\end{equation}
We define
\begin{equation}
\begin{aligned}
  \mathcal G_s(\ell,m)   &= \int \dd^2\hat{\bm K} Y_1^1(\hat{\bm K}) Y_\ell^m(\hat{\bm K}) Y_{\ell+s}^{-(m+1)}(\hat{\bm K}).
  \end{aligned}
\end{equation}
Equating the two  expressions finally  gives
\begin{equation}
\begin{aligned}
	&A_{\ell m} \left(\partial_a - \frac{m}{a}\right)j_\ell(Ka)= \\ & -\ii K\sqrt{\frac{8\pi}{3}} \sum_{s=\pm1} \ii^s(-1)^{m+1} A_{\ell+s,m+1} \mathcal G_s(\ell,m)  j_{\ell+s}(Ka).
\end{aligned}
  \label{eq:app_recoupling_identity}
\end{equation}
It is useful to define
\begin{equation}
\mathcal{H}_s(\ell,m) \equiv (-1)^{m+1} \mathcal{G}_s(\ell,m)
\end{equation}
which can be written in a simple closed form using spherical harmonics orthogonality and recurrence relations:
\begin{equation}
\mathcal{H}_s(\ell,m)=  s\left[  \frac{  3\left(\ell+s m+\frac{1+s}{2}\right) \left(\ell+s m+\frac{1+3s}{2}\right)}{8\pi(2\ell+1)(2\ell+1+2s)} \right]^{1/2},  \quad s=\pm 1.
  \label{eq:Hs_single_closed}
\end{equation}
Substituting Eq.~\eqref{eq:app_recoupling_identity} into Eq.~\eqref{eq:Fplus_uni_bessel_integral_new} gives
\begin{equation}
\mathcal  F_+^{\rm uni} = \widetilde{K}_0 \sum_{\ell,m}\sum_{s=\pm1} \ii^s A_{\ell m}A_{\ell+s,m+1} \mathcal{H}_s(\ell,m) \mathcal R_{\ell,\ell+s}(m),
\end{equation}
where
\begin{equation}
  \widetilde K_0=-\frac{(4\pi)^4\ii}{(2\pi)^3}\sqrt{\frac{4\pi}{3}}\frac{G^2M^2\rho_0}{c_s^2},
\end{equation}
and all remaining radial dependence is contained in
\begin{equation}
  \mathcal R_{\ell,\ell+s}(m) = \int_0^\infty \dd z \frac{z\,j_\ell(z)j_{\ell+s}(z)}{z^2-(m\M+\ii\epsilon)^2},
  \quad \M=\frac{\Omega a}{c_s}.
\end{equation}
For $m>0$ \citep[][]{Desjacques2022},
\begin{equation}
  \mathcal R_{\ell,\ell+s}(m)=  \frac{\ii\pi}{2} j_{L}(m\M)h_{L-1}^{(1)}(m\M),
  \label{eq:app_R_positive}
\end{equation}
where $L=\max(\ell,\ell+s)$. The negative and zero-frequency values are respectively
\begin{equation}
  \mathcal R_{\ell,\ell+s}(-m)  =  \mathcal R_{\ell,\ell+s}(m)^*,  \quad m>0,
\end{equation}
and
\begin{equation}
  \mathcal R_{\ell,\ell+s}(0)  =  \frac{\pi}{2(4L^2-1)}.
\end{equation}

\section{Analytic power-law Green function}
\label{sec:anal_green}

Here we derive the radial acoustic Green function for power-law density stratification with constant background sound speed case, Eq.~\eqref{eq:power_law_profile}. We start from the homogeneous radial equation
\begin{equation}
  \left[\mathcal L_\ell-\omega^2\frac{\rho_0}{c_s^2}\right] u_\ell = 0 ,
  \label{eq:homo_rad}
\end{equation}
where $u_\ell$ is a basis function used to build the Green function. Using the expression of the radial operator Eq.~\eqref{eq:Lell_def}, Eq.~\eqref{eq:homo_rad} becomes
\begin{equation}
  u_\ell''+\frac{p+2}{r}u_\ell' +\left[k^2-\frac{\ell(\ell+1)}{r^2}\right]u_\ell=0 .
  \label{eq:power_law_radial_eq}
\end{equation}
Let 
\begin{equation}
  x\equiv kr, \quad u_\ell(r)=r^{-\mu}Z(x) ,
\end{equation}
with $\mu \equiv (p+1) /2$. After substitution, Eq.~\eqref{eq:power_law_radial_eq} becomes the standard Bessel equation 
\begin{equation}
	x^2Z''+xZ'+\left(x^2-\nu_{\ell p}^2\right)Z=0,
\end{equation}
with
\begin{equation}
\nu_{\ell p}^2\equiv \ell(\ell+1)+\mu^2 .
\end{equation}
The two  basis functions are therefore:
\begin{align}
  u_{\ell p}^{J}(r;k)
  &=\left(\frac{\pi}{2k}\right)^{1/2}r^{-\mu}J_{\nu_{\ell p}}(kr),
  \label{eq:power_law_uJ}\\
  u_{\ell p}^{H}(r;k)
  &=\left(\frac{\pi}{2k}\right)^{1/2}r^{-\mu}H^{(1)}_{\nu_{\ell p}}(kr),
  \label{eq:power_law_uH}
\end{align}
where the prefactor $(\pi/2k)^{1/2}$ is chosen such that we recover the homogeneous medium Bessel-Hankel Green function Eq.~\eqref{eq:G_uniform_ansatz_new} when $p=0$. Equation~\eqref{eq:power_law_uJ} is the solution that is regular at the origin and Eq.~\eqref{eq:power_law_uH} is the outgoing solution ($H_\nu^{(1)}(kr)\sim \exp(\ii kr)$) at large positive $r$ \citep{Haber2024}. The Green function can be written in the product form
\begin{equation}
  g_\ell^{(p)}(r,r';\omega)=C_{\ell p}(k)  u_{\ell p}^{J}(r_<;k)u_{\ell p}^{H}(r_>;k) ,
\end{equation}
where $C_{\ell p}(k)$ is obtained by integrating Eq.~\eqref{eq:Green_strat} from $r = r'-\epsilon$ to $r = r' + \epsilon$ where $\epsilon$ is a positive infinitesimal quantity \citep{Haber2024}, yielding
\begin{equation}
C_{\ell p}(k) = {\ii k a^p \over \rho_a} .
\end{equation}

\section{Asymptotic Mach-number limits for the constant sound speed power-law case}
\label{app:asymptotic}

Here we derive the low- and high-Mach asymptotic limits of the dynamical friction force for power-law density stratification with constant background sound speed case, Eq.~\eqref{eq:power_law_profile}. 
\subsection{Low-Mach limit}
When $\M \ll 1$, $\Omega \ll c_a /a$, that is the response is quasi-static to leading order $\nabla (h + \Phi_p) = 0$, therefore
\begin{equation}
  h^{(0)}(\xx) = {GM\over a} \sum_{\ell=1}^{\infty}\sum_{m=-\ell}^{\ell}{4\pi\over 2\ell+1} Y_{\ell m}(\hat{\XX})Y_{\ell m}^{*}(\xhat)Q_\ell(x)+C, 
\end{equation}
where $C$ is the monopole part and we have introduced $x = r/a$, $X = R/a$ and $Q_\ell = x^\ell_</ x^{\ell+1}_>$. For now on, we omit the $(0)$ superscript and assume perturbations to be constituted of their leading order only. Using $\rho_0 = \rho_a x^p$ yields
\begin{equation}
  \rho_{1,\ell m}(x) = {GM\rho_a\over a c_a^2}{4\pi\over 2\ell+1} Y_{\ell m}^{*}(\xhat) x^p Q_\ell(x).
  \label{eq:rho_lm_coeff}
\end{equation}
Inserting Eq.~\eqref{eq:rho_lm_coeff} into Eq.~\eqref{eq:wake_potential_def} yields
\begin{equation}
\begin{aligned}
  \Psi_{\ell m}(\RR) &= -{G^2M\rho_a a\over c_a^2} \left({4\pi\over 2\ell+1}\right)^2 Y_{\ell m}^{*}(\xhat)Y_{\ell m}(\hat\RR) \\ 
		     & \times \int_0^\infty x^{p+2}Q_\ell(x)Q_\ell(x;X) \dd x,
\end{aligned}
\end{equation}
in the limit $\omega \to 0$. The radial force on the perturber is 
\begin{equation}
  F_r=-M{\partial\Psi\over\partial R}\bigg|_{R=a,\hat\RR=\xhat}.
\end{equation}
Using
\begin{equation}
  {\partial Q_\ell(x;X)\over\partial X}\bigg|_{X=1}=
  \begin{cases}
      -(\ell+1)x^\ell, & x<1,\\
      (\ell+1)x^\ell, & x>1,
  \end{cases}
\end{equation}
the radial integral appearing in the force thus reads
\begin{equation}
\begin{aligned}
B_\ell(p) &= -\int_0^\infty x^{p+2}Q_\ell(x){\partial Q_\ell(x;X)\over\partial X}\bigg|_{X=1}\dd x \\
	  &=  (\ell+1)\int_0^1 x^{p+2\ell+2}\dd x - \ell \int_1^\infty x^{p-2\ell}\dd x .
\end{aligned}
  \label{eq:rad_integrale}
\end{equation}
For Eq.~\ref{eq:rad_integrale} to converge at infinity one needs $2\ell -p -1 > 0$. We find
\begin{equation}
  B_\ell(p) = {\ell+1\over p+2\ell+3} - {\ell\over 2\ell-p-1}.
\end{equation}
Finally, using Unsöld's theorem 
\begin{equation}
	\sum_{m=-\ell}^\ell Y_{\ell m}^\ast(\xhat) Y_{\ell m}(\xhat) =  {2\ell + 1 \over 4\pi},
\end{equation}
and using the zero-inner monopole gauge removing the inner monopole yields
\begin{equation}
	\Re\left[I^{\rm low}(p;\ell_{\max})\right]  = \M^2 \sum_{\ell=1}^{\ell_{\max}} {1\over 2\ell+1}\left({\ell+1\over p+2\ell+3} - {\ell\over 2\ell-p-1}\right).
\end{equation}
The azimuthal force vanishes in the strictly static low-Mach limit:
\begin{equation}
	\Im\left[I^{\rm low}(p;\ell_{\max})\right] = 0
\end{equation}
that is because the static wake is symmetric under reflection through the line joining the origin and the perturber. There is no preferred forward or backward direction along the orbit. 
In the $p=0$ case, we do not use a zero-inner-monopole convention gauge, but instead enforce decaying enthalpy perturbation at infinity. We thus cannot omit the monopole term $B_0(0) = 1 / 3$. 

\subsection{High-Mach limit}
In the high-Mach limit and for fixed $\ell$, the wave operator $\mathcal L_\ell - (m\M/a)^2$ contains terms of order $O(\M^2)$. Hence, the response of that fixed mode is suppressed as $O(\M^{-2})$. The only mode that is not suppressed is $m=0$. The radial integral appearing in the force is still Eq.~\eqref{eq:rad_integrale}, but the angular factor becomes
\begin{equation}
Y_{\ell 0}^\ast(\hat\RR) Y_{\ell 0}(\hat\RR) = {2 \ell + 1 \over 4\pi} P_\ell(0)^2.
\end{equation}
Finally, we get
\begin{equation}
	\Re\left[I^{\rm high}(p;\ell_{\max})\right]  = \M^2 \sum_{\ell=1}^{\ell_{\max}} {P_\ell(0)^2 \over 2\ell+1}\left({\ell+1\over p+2\ell+3} - {\ell\over 2\ell-p-1}\right).
\label{eq:Ir_low_general}
\end{equation}
Because only the axisymmetric mode $m=0$ survives in the $\M \to \infty$ limit, 
\begin{equation}
	\Im\left[I^{\rm high}(p;\ell_{\max})\right] = 0.
\end{equation}
As for the low-Mach limit we cannot omit the monopole term $B_0(p) = 1 / (3 + p)$ for the $p=0$ case because we enforce decaying enthalpy perturbation at infinity. A general expression for the asymptotic low- and high-Mach limit reads
\begin{equation}
	\Re\left[I^{\rm asy}\right] = {\chi_0 \over p+3} + \sum_{\ell=1}^{\ell_{\max}} {\mathcal W_\ell \over 2\ell + 1 } \left({\ell+1\over p+2\ell+3} - {\ell\over 2\ell-p-1}\right),
\end{equation}	
where $\chi_0 = 0$ for the zero-inner-monopole convention and $\chi_0 = 1$ for the full static response, and 
\begin{equation}
	\mathcal W_\ell =
  \begin{cases}
      1, & \M \ll 1 ,\\
      P_\ell(0)^2, & \M \gg 1.
  \end{cases}
  \label{eq:w_ell}
\end{equation}
\section{Extension to double-perturber systems}
\label{app:double}
Let
\begin{equation}
  M_2=qM_1, \quad a_1=\frac{q}{1+q}\ab,  \quad a_2=\frac{1}{1+q}\ab,
\end{equation}
where $a$ is now the binary separation and $q = M_2 / M_1$ is the binary mass ratio.  We write the two circular trajectories as
\begin{equation}
  \rr_j(t)=a_j\left[\cos(\Omega t+\alpha_j)\xhat + \sin(\Omega t+\alpha_j)\yhat\right],
  \quad j=1,2,
\end{equation}
with $\alpha_1 = 0$, and $\alpha_2 = \pi$. The double perturber gravitational potential now reads
\begin{equation}
	\Phi_p (\bx, t) = - \sum_{j=1}^{2} \frac{GM_j}{\| \bx - \rr_j(t) \|} ,
\end{equation}
and the radial amplitude of the perturber potential, excluding the factor $e^{-\ii m\Omega t}$, Eq.~\eqref{eq:Phi0_q_def}, becomes
\begin{equation}
  \Phi_{\ell m}^{(0)}(r)
  =-GM_1\frac{4\pi}{2\ell+1}A_{\ell m}
  \left[q_\ell(r,a_1)+q(-1)^m q_\ell(r,a_2)\right].
\end{equation}
It is useful to distinguish the radius $a_A$ at which the wake is originated from the radius $a_B$ at which the force is evaluated. We define
\begin{equation}
\begin{aligned}
\mathfrak F_{\ell m}(a_{B},a_A)
  &\equiv   \int_0^\infty \dd r\, r^2 w(r) \left[\partial_R q_\ell(r,R)-\frac{m}{R}q_\ell(r,R)\right]_{R=a_{B}} \\
  & \times
  \int_0^\infty \dd r'r'^2 g_\ell(r,r'; m\Omega)\mathcal L_\ell q_\ell(r',a_A) .
  \label{eq:Frak_lm_two_radius_def}
\end{aligned}
\end{equation}
The helicity force on body $B$ can then be written as
\begin{equation}
	F_{B,+}(t)=\ee^{\ii(\Omega t+\alpha_{B})}\mathcal F_{B,+},
\end{equation}
where
\begin{equation}
  \mathcal F_{B,+} =\frac{G^2M_B}{\sqrt2} \sum_{\ell=0}^{\infty}\sum_{m=-\ell}^{\ell} \left(\frac{4\pi}{2\ell+1}\right)^2 A_{\ell m}^2
  \sum_{A=1}^{2}M_A\ee^{\ii m(\alpha_{B}-\alpha_A)}\mathfrak F_{\ell m}(a_B,a_A) .
  \label{eq:FBplus_binary_general}
\end{equation}
This reduces to Eq.~\eqref{eq:Fplus_strat_matrix} when $a_A=a_B=a$. For $\alpha_1=0$, and $\alpha_2=\pi$,  Eq.~\eqref{eq:FBplus_binary_general} gives
\begin{equation}
\begin{aligned}
  \mathcal F_{1,+} &=\frac{G^2M_1^2}{\sqrt2} \sum_{\ell m} \left(\frac{4\pi}{2\ell+1}\right)^2 A_{\ell m}^2 \\
		   & \times \left[ \mathfrak F_{\ell m}(a_1,a_1) +q(-1)^m\mathfrak F_{\ell m}(a_1,a_2) \right],
  \label{eq:F1plus_binary_q}\\
\end{aligned}
\end{equation}
\begin{equation}
\begin{aligned}
  \mathcal F_{2,+} &=\frac{G^2M_1^2}{\sqrt2}q \sum_{\ell m} \left(\frac{4\pi}{2\ell+1}\right)^2 A_{\ell m}^2 \\ 
		   & \times \left[ (-1)^m\mathfrak F_{\ell m}(a_2,a_1) +q\mathfrak F_{\ell m}(a_2,a_2) \right].
\end{aligned}
\end{equation}
The terms with $A=B$ are the forces from each body's own wake, while the terms with $A\ne B$ are cross-wake forces. Finally, the physical components of the force acting on body $B$ are
\begin{equation}
  F_{B,r}=\sqrt2\Re(\mathcal F_{B,+}), \quad  F_{B,\varphi}=\sqrt2\Im(\mathcal F_{B,+}) .
\end{equation}
We now consider the uniform limit in exactly the same manner as in the single-perturber calculation.  The only difference is that the source radius and the force-evaluation radius are kept distinct. Eq.~\eqref{eq:Lq_uniform_delta_new}, now with the source radius $a_A$, reads
\begin{equation}
  \mathcal L_\ell q_\ell(r,a_A) =\rho_0\frac{2\ell+1}{a_A^2}\delta(r-a_A).
\end{equation}
The radial integral in Eq.~\eqref{eq:Frak_lm_two_radius_def} thus simply becomes
\begin{equation}
  \int_0^\infty \dd r'r'^2 g_\ell(r,r') \mathcal L_\ell q_\ell(r',a_A) =\rho_0(2\ell+1)g_\ell(r,a_A; m\Omega) .
\end{equation}
Inserting the uniform retarded Green function, Eq.~\eqref{eq:G_uniform_ansatz_new}, yields
\begin{equation}
  \rho_0(2\ell+1)g_\ell(r,a_A) =(2\ell+1)\ii k j_\ell(kr_<^A)h_\ell^{(1)}(kr_>^A),
\end{equation}
where
\begin{equation}
  r_<^A\equiv \min(r,a_A), \quad r_>^A\equiv \max(r,a_A),
\end{equation}
Substitution into Eq.~\eqref{eq:Frak_lm_two_radius_def} gives the analogue of Eq.~\eqref{eq:Fplus_uni_from_strat_new}, but with a wake launched at $a_A$ and a force kernel evaluated at $a_B$:
\begin{equation}
\begin{aligned}
  \mathfrak F^{\rm uni}_{\ell m}(a_B,a_A) &=\frac{\rho_0}{c_s^2}(2\ell+1)\ii k_m \int_0^\infty\dd r\, r^2 \\
					  & \times \left[\partial_Rq_\ell(r,R)-\frac{m}{R}q_\ell(r,R)\right]_{R=a_B} j_\ell(k_m r_<^A)h_\ell^{(1)}(k_m r_>^A) .
  \label{eq:Frak_uni_two_radius_hankel_step}
\end{aligned}
\end{equation}
Using the same Fourier--Bessel representation as in Eq.~\eqref{eq:q_bessel_identity_new}, but applying the derivative at the evaluation radius yields
\begin{equation}
\begin{aligned}
  \left[\partial_R-\frac{m}{R}\right]_{R=a_B}q_\ell(r,R)  &=\frac{2(2\ell+1)}{\pi} \\ 
							  & \times \int_0^\infty\dd Kj_\ell(Kr) \left[\partial_{a_B}-\frac{m}{a_B}\right]j_\ell(Ka_B).
\end{aligned}
\end{equation}
The remaining $r$-integral is the same Green-function identity as Eq.~\eqref{eq:bessel_green_identity_new}, replacing $a$ by the source radius $a_A$:
\begin{equation}
  \int_0^\infty\dd r\, r^2j_\ell(Kr) \ii k_m j_\ell(k_m r_<^A)h_\ell^{(1)}(k_m r_>^A) =\frac{j_\ell(Ka_A)}{K^2-(k_m+\ii\epsilon)^2} .
  \label{eq:bessel_green_identity_binary}
\end{equation}
Combining Eqs.~\eqref{eq:Frak_uni_two_radius_hankel_step}--\eqref{eq:bessel_green_identity_binary} therefore gives
\begin{equation}
\begin{aligned}
  \mathfrak F^{\rm uni}_{\ell m}(a_B,a_A)  &=\frac{2\rho_0}{\pi c_s^2}(2\ell+1)^2 \\
					   & \times \int_0^\infty \dd K \frac{j_\ell(Ka_A)}{K^2-(k_m+\ii\epsilon)^2} \left[\partial_{a_B}-\frac{m}{a_B}\right]j_\ell(Ka_B).
\end{aligned}
\label{eq:Frak_uni_two_radius}
\end{equation}
Multiplying Eq.~\eqref{eq:Frak_uni_two_radius} by $A_{\ell m}^2$ and using Eq.~\eqref{eq:app_recoupling_identity}, gives
\begin{equation}
\begin{aligned}
  A_{\ell m}^2\mathfrak F^{\rm uni}_{\ell m}(a_B,a_A) &=\frac{2\rho_0}{\pi c_s^2}(2\ell+1)^2 \left[-\ii\sqrt{\frac{8\pi}{3}}\right] \\
						      & \times \sum_{s=\pm1}\ii^s A_{\ell m}A_{\ell+s,m+1}{\mathcal H}_s(\ell,m) \R_{\ell,\ell+s}^{BA}(m)
  \label{eq:Frak_uni_two_radius_after_recoupling_m}
\end{aligned}
\end{equation}
where
\begin{equation}
  \R_{\ell,\ell+s}^{BA}(m) \equiv \int_0^\infty\dd K \frac{Kj_\ell(Ka_A)j_{\ell+s}(Ka_B)}{K^2-(k_m+\ii\epsilon)^2},
\end{equation}
and \citep{Desjacques2022}
\begin{equation}
  \mathcal R_{\ell,\ell-1}^{BA}(m)
  =
  \begin{cases}
  \displaystyle
  \frac{\ii\pi}{2} \left[ j_{\ell-1}(k_m a_B) h_\ell^{(1)}(k_m a_A) - \frac{a_B^{\ell-1}}{a_A^{\ell+1}}\frac{1}{k_m^2} \right],  & a_A>a_B,
  \\
  \displaystyle
  \frac{\ii\pi}{2} j_\ell(k_m a_A) h_{\ell-1}^{(1)}(k_m a_B),  & a_A<a_B .
  \end{cases}
\end{equation}
The value at $a_A=a_B$ reduces to Eq.~\eqref{eq:app_R_positive}. The superscript $BA$ means that the wake is generated at $a_A$ and the force is
\begin{equation}
	\begin{aligned}
  \mathcal F_{B,+}^{\rm uni} &=\frac{G^2M_B}{\sqrt2}  \sum_{A=1}^{2}M_A \sum_{\ell m} \left(\frac{4\pi}{2\ell+1}\right)^2 \ee^{\ii m(\alpha_B-\alpha_A)} \\
   & \times \mathcal K_\ell \sum_{s=\pm1} (-\ii)\ii^s  A_{\ell m}A_{\ell+s,m+1} \mathcal H_s(\ell,m) \R_{\ell,\ell+s}^{BA}(m),
\end{aligned}
\end{equation}
where
\begin{equation}
  \mathcal K_\ell \equiv \frac{2\rho_0}{\pi c_s^2}(2\ell+1)^2 \sqrt{\frac{8\pi}{3}} .
\end{equation}

\section{Extra figures}
\begin{figure}
\resizebox{\hsize}{!}{\includegraphics{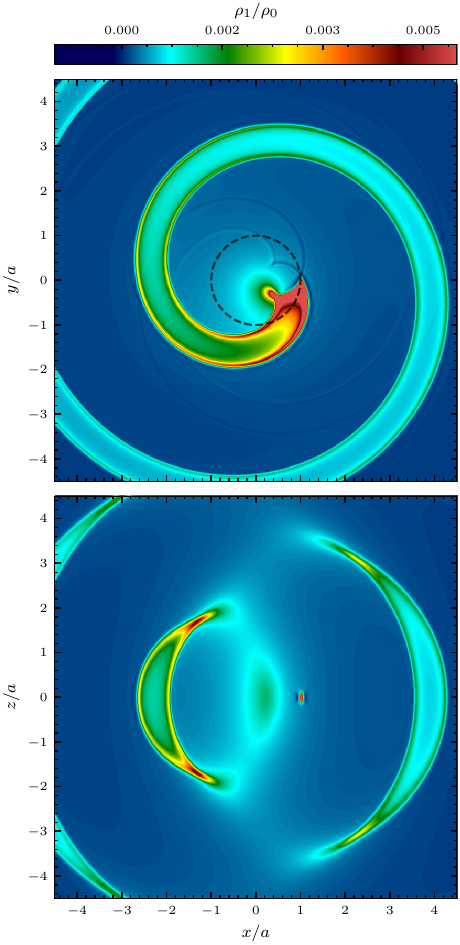}}
\caption{Steady-state acoustic wake overdensity $\rho_1/\rho_0$ on the $xy$ (top) and $xz$ (bottom) planes for a homogeneous medium, $\M = 2$ and $\ell_{\rm max} = 96$. The perturber is located at $(x_p,y_p,z_p) = (a,0,0)$ and has mass $M= 10^{-3} c_s^2 a/G$. The black circle indicates its circular orbit.}
\label{fig:wake_uniform}
\end{figure}
\begin{figure}[t]
    \resizebox{\hsize}{!}{
    \includegraphics[]{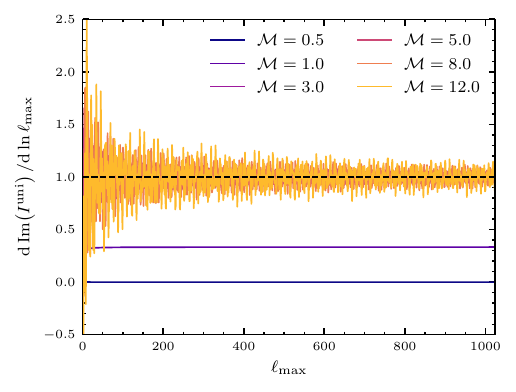}}
    \caption{$\dd \Im\!\left(I^{\rm uni}\right)/\dd\ln\ell_{\rm max}$ as a function of the truncated maximum harmonic degrees $\ell_{\rm max}$ for various Mach numbers $\M$. $\dd \Im\!\left(I^{\rm uni}\right)/\dd\ln\ell_{\rm max} \to 1$ for all $\M > 1$, indicating a Coulomb logarithm scaling of the drag force in the linear theory for the supersonic circular motion of the perturber.}
    \label{fig:logI_ell}
\end{figure}
\begin{figure}
  \resizebox{\hsize}{!}{\includegraphics{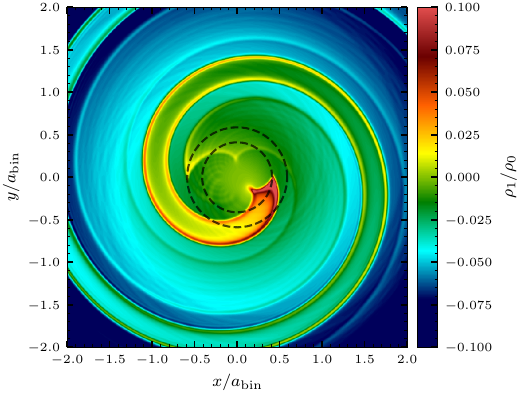}}
  \caption{Same as Fig.~\ref{fig:wake_q1_pm2}, but for a perturber mass-ratio $q=0.7$.}
  \label{fig:wake_q07_pm2}
\end{figure}
\begin{figure}
  \resizebox{\hsize}{!}{\includegraphics{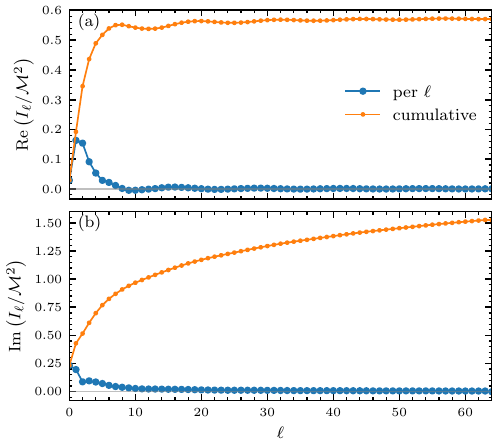}}
  \caption{Cumulative and individual harmonic contribution to $\Re\!\left(I\right)/\M^2$ (panel (a)) and $\Im\!\left(I\right)/\M^2$ (panel (b)) defined in Eq.~\eqref{eq:normalized_force} for a perturber on a circular orbit with radius $a = R_\ast/2$ ($\M \simeq 1.75$) in the RG MESA model.}
\label{fig:F_ell_RGt}
\end{figure}
\begin{figure}
  \resizebox{\hsize}{!}{\includegraphics{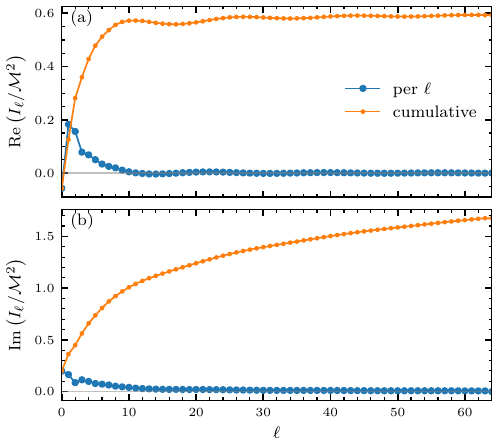}}
  \caption{Same as Fig.~\ref{fig:F_ell_RGt}, but for the RSG MESA model ($\M \simeq 1.60$).}
\label{fig:F_ell_RSG}
\end{figure}
\begin{figure}
  \resizebox{\hsize}{!}{\includegraphics{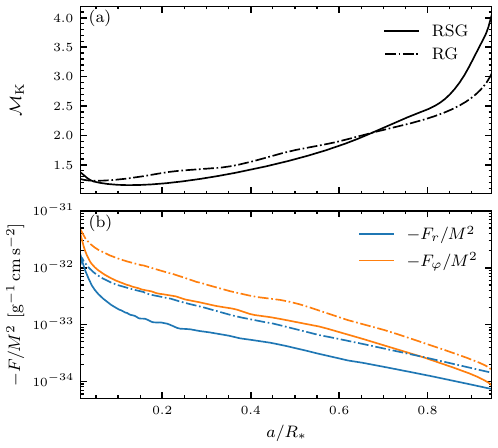}}
  \caption{Keplerian orbital Mach number  $\M_{\rm K}$ (panel (a)) and radial and azimuthal forces exerted on the perturber (panel (b)) as a function of the perturber's orbital radius within the RSG and RG envelopes.}
  \label{fig:mesa_force_vs_a}
\end{figure}
\begin{figure}
  \resizebox{\hsize}{!}{\includegraphics{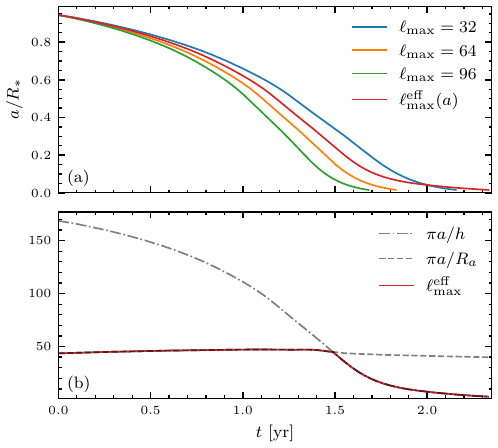}}
  \caption{Orbital decay of a $0.03\, M_\odot$ perturber in a RG MESA model obtained by integrating Eq.~\eqref{eq:mesa_adot} along a sequence of circular Keplerian orbits for different truncation harmonic degrees,  and for $\ell_{\rm max}^{\rm eff}(a)$ (shown in panel (b)). }
  \label{fig:a_vs_time_diff_lmax}
\end{figure}
\begin{figure}[t]
    \resizebox{\hsize}{!}{
    \includegraphics[]{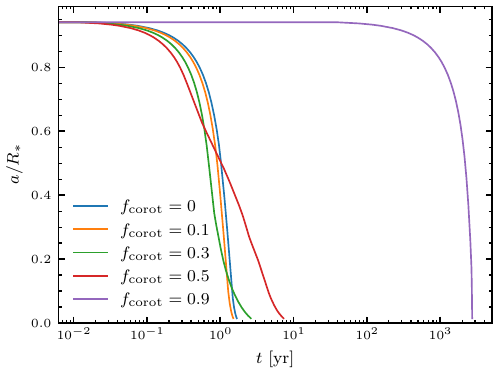}}
    \caption{Orbital decay of a $0.03\, M_\odot$ perturber in a RG MESA model obtained by integrating Eq.~\eqref{eq:mesa_adot} along a sequence of circular Keplerian orbits for different degrees of corotation of the envelope $f_{\rm corot}$, and for $\ell_{\rm max}= 96$.}
    \label{fig:adot_fcorot}
\end{figure}
\begin{figure}
  \resizebox{\hsize}{!}{\includegraphics{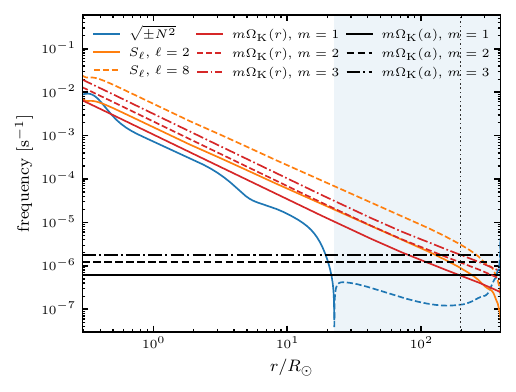}}
  \caption{Same as Fig.~\ref{fig:mesa_frequencies_RGt}, but for a $0.1~M_\odot$ perturber located at $a=0.5\,R_\ast$ in the RSG MESA stellar model.}
  \label{fig:mesa_frequencies_RSG}
\end{figure}
\end{appendix}
\end{document}